\documentclass[aps,twocolumn,showpacs,pra,amsmath,amssymb,floatfix,superscriptaddress]{revtex4-2}
\usepackage{amsmath,amssymb,graphicx,color}
\usepackage{bm}
\usepackage{bbm}
\usepackage[caption=false]{subfig}  
\usepackage[usenames,dvipsnames]{xcolor}
\usepackage[normalem]{ulem}
\usepackage{soul}
\usepackage[colorlinks=true,citecolor=Cerulean,linkcolor=RubineRed,urlcolor=Cerulean]{hyperref}
\usepackage{soul,xcolor}

\newcommand{\Tr}{\ensuremath{\mathrm{Tr}\,}}

\usepackage{mathptmx}
\DeclareMathAlphabet{\mathcal}{OMS}{cmsy}{m}{n}
\DeclareSymbolFont{largesymbols}{OMX}{cmex}{m}{n}

\usepackage{braket}

\usepackage{verbatim} 

\begin{document}
\title{Nonadiabatic evolution and thermodynamics for a boundary-driven system \\ with a weak intrasubsystem interaction}

\author{Chao Jiang} \email{cjiang@gscaep.ac.cn}
\affiliation{Graduate School of China Academy of Engineering Physics, Beijing 100193, China}

\author{Lei Shao}\email{lshao@gscaep.ac.cn}
\affiliation{Graduate School of China Academy of Engineering Physics, Beijing 100193, China}

\begin{abstract}

We derive a time-dependent master equation for an externally driven system, {where the} subsystems weakly interact with each other and {are} locally {connected} to the thermal reservoirs. The nonadiabatic equation obtained here can be viewed as a generalization of the local master equation, which has already been extensively used in describing the dynamics of a boundary-driven system. In addition, we investigate the fundamental reason underlying the thermodynamic inconsistency generated by the local and nonadiabatic master equations. {We also provide a criterion to assess the consistency of these two equations with the second law of thermodynamics.} Finally, we numerically confirm our results by considering a toy model consisting of two qubits and two local heat baths.

\end{abstract}

\maketitle

\section{Introduction \label{sec:intro}}  

The research of a many-body system coupling to one or multiple reservoirs at its edge is an extremely significant topic in the nonequilibrium thermodynamics. The system referred is usually named as boundary-driven system in the literature \cite{RevModPhys.94.045006}. The study on the boundary-driven system dates back to the Fourier heat transfer law in which a piece of material is placed between the hot and cold heat baths \cite{RevModPhys.94.045006, Fourier}. Recent decades have witnessed considerable progress in the investigation of the boundary-driven system, such as manipulating the dynamics of a quantum system \cite{Popkov_2013, schindler2013quantum, PhysRevLett.111.040601, PhysRevE.94.042135}, exploring the novel phenomena in an open system far from equilibrium \cite{PhysRevLett.101.105701, Mitchison_2018, PhysRevLett.105.060603}, and constructing the quantum thermal machines \cite{PhysRevLett.105.130401, PhysRevE.89.032115, PhysRevLett.108.120602}.

The dynamics of an open quantum system is typically described by the master equation \cite{weiss2012quantum, HMW, KJ, HPBFP}. {The conventional derivation of the master equation is based on the global approach, in which the system-bath interaction is decomposed according to the eigenvectors of the whole system's Hamiltonian, and the equation obtained accordingly is called the global master equation \cite{RevModPhys.94.045006, Levy_2014, Hofer_2017, PhysRevA.93.062114, gonzalez2017testing, Cattaneo_2019}. However, the derivation becomes quite involved for the boundary-driven system, especially when the form of the intersubsystem interaction is intricate and the number of subsystems is large \cite{Trushechkin_2016}. As an alternative, it is more intuitive to obtain the local master equation (LME) \cite{RevModPhys.94.045006, Levy_2014} for such kind of system based on the local approach, where only the knowledge of the energy spectrum of each individual subsystem is required. Due to its lower computational complexity, the LME has received a broad employment in describing the energy transfer of a boundary-driven system \cite{PhysRevLett.105.130401, PhysRevE.107.034128, PhysRevE.94.042122, PhysRevLett.107.137201, PhysRevE.107.044102, PhysRevA.89.022128}.}

Although the LME is an efficient tool to model the open quantum system, it is only applicable to the situation where the Hamiltonian of the system is time-independent. Nevertheless, for quite a lot attractive and important research projects in nonequilibrium statistical physics, the system of interest is influenced by not only the reservoir's perturbation but also the external driving \cite{curzon1975efficiency, PhysRevB.98.064307, PhysRevB.92.014306, PhysRevE.92.032113, PhysRevA.104.032201}. Therefore, {it is necessary and urgent to develop a more general LME applicable for the boundary-driven system with a time-dependent Hamiltonian.}

Apart from its limited application scope, the LME has been found to {potentially contradict the second law of thermodynamics} \cite{Levy_2014, vcapek2002zeroth, stockburger2017thermodynamic}. Due to its importance in characterizing the dynamics of the boundary-driven system, it has inevitably become a crucial topic to investigate the reasons and specific circumstances under which the LME violates this fundamental thermodynamic principle. Even though considerable efforts have been made, the existing works have mainly focused on
reconciling the LME with the second thermodynamic law, such as adopting the alternative global approach \cite{Levy_2014, Trushechkin_2016}, redefining the expression of thermodynamic quantities \cite{barra2015thermodynamic, De_Chiara_2018, RevModPhys.93.035008, PhysRevResearch.3.013165}, and modifying the form of the LME by using an additional secular approximation \cite{PhysRevE.107.014108}. {There has been little theoretical analysis on why the original equation leads to thermodynamic contradictions, and this remains a confusing issue.}

In this text, we deal with these two problems. {The first objective of the paper is to generalize the standard LME and to derive a time-dependent local master equation (TDLME), which is applicable for the boundary-driven system subjected to the external driving. The second objective is to reveal the fundamental reason for the thermodynamic inconsistency generated by the LME, and to establish the valid range within which the LME remains consistent with the second thermodynamic law.
	
Inspired by the methods developed in Refs.~\cite{PhysRevE.107.014108, PhysRevA.104.0322012021}, we start our derivation in the adiabatic frame of reference. Next, we apply a particularly crucial iterative technique employed in Ref.~\cite{RevModPhys.94.045006} to reformulate the quantum Liouville equation. Following the standard treatment, we then adopt the Born–Markov approximation \cite{HMW, KJ, HPBFP} to obtain the Redfield-like equation. However, in contrast to the standard Redfield equation \cite{RevModPhys.94.045006}, here both the phase factors and the jump operators in the dissipation terms are complicated functions of time. By assuming that the bath correlation time is short enough, these time-dependent terms can be approximated by their respective first-order Taylor expansions. Finally, we reduce the TDLME into a more conveniently employed form by performing the secular rotating wave approximation \cite{HMW, HPBFP}. The coefficients in the equation are analytically worked out later by considering the bosonic heat bath with an Ohmic spectral density. }

Furthermore, {we also present a theoretical explanation for the thermodynamic inconsistency in the LME and TDLME.} We find that the heat currents near the steady state calculated by these two equations are third order small quantities. However, the LME and the TDLME themselves are only kept up to the second order small quantity. Therefore, the LME and the TDLME are inherently unsuitable to describe the evolution close to the steady state, and it is intuitive that these two equations can introduce the thermodynamic contradiction in this region. {In contrast, when considering the equations within their appropriate range of application, where the thermodynamic observables evaluated accordingly are at least second-order small quantities, we find that the violation of the second thermodynamic law can be avoided \cite{RevModPhys.94.045006, PhysRevResearch.3.013165}.} Our results are numerically verified by a toy model which contains a pair of time-dependent boundary-driven qubits.

The structure of the article is as follows. In Sec.~\ref{sec:derivation}, we give a theoretical derivation for the TDLME in the most general case. In Sec.~\ref{sec:bosonic}, we specify the nature of the local heat bath as well as the structure of the system-bath interaction, and exactly work out the form of the TDLME. In Sec.~\ref{sec:condition}, we discuss the reason which leads to the thermodynamic inconsistency in the LME and the TDLME. The valid range of two equations is also established after that. Section \ref{sec:example} presents a numerical confirmation of our conclusions and Sec.~\ref{sec:conclusion} summarizes the paper { and provides a future prospect.}

\section{TDLME for a boundary-driven system subjected to the external driving}
\label{sec:tdlme}

{\subsection{Microscopic derivation for the TDLME}
\label{sec:derivation}
}

The system of interest consists of $N$ subsystems, which are denoted by $s_1$, $s_2$, $\cdots$, $s_N$. The interaction between these subsystems is assumed to be weak, and each subsystem is weakly connected to a local Markovian heat bath $B_i$, which is prepared in a thermal  equilibrium state $\rho_{B_i}$ with temperature $T_i$. The full Hamiltonian of the whole composite system, i.e., the system under consideration and the thermal reservoirs, takes the form \cite{RevModPhys.94.045006, PhysRevE.107.014108}
\begin{align}
H(t) & = H_0(t) +  \lambda H_I  + H_B + \zeta V
\notag
\\
& = \sum_{i = 1}^{N} H_{s_i}(t) + \lambda H_I  + \sum_{i = 1}^{N} H_{B_i} + \zeta \sum_{i = 1}^{N} V_i.
\end{align}
Here, $\lambda$ and $\zeta$ are two dimensionless  expansion parameters that can be viewed as small quantities during the theoretical derivation but eventually can be set to be unity \cite{griffiths2018introduction}. $H_{s_i}(t)$ denotes the bare Hamiltonian of the subsystem $s_i$, i.e., Hamiltonian without the intrasubsystem interaction, and $H_0(t) = \sum_{i = 1}^{N} H_{s_i}(t)$ is the bare Hamiltonian of all the subsystems concerned. The change of $H_0(t)$ within the time scale of the bath correlation time $\tau_B$ is assumed to be small so that the nonadiabatic transition between different energy levels of $H_0(t)$ over $\tau_B$ is weak. $H_I$ denotes the time-independent interaction between subsystems, $ H_{B_i}$ represents the Hamiltonian of the local heat bath $B_i$, and $V_i$ is the coupling between the subsystem $s_i$ and its corresponding local thermal reservoir $B_i$. Without loss of the generality, $\lambda H_I$ and $\zeta V_i$ are supposed to be of the similar magnitude and both of them are weak enough such that they can be treated as perturbations.

Combining the methods proposed in Refs.~\cite{PhysRevE.107.014108, PhysRevA.104.0322012021}, we introduce the following adiabatic frame of reference transformation operator,
\begin{align}
U(t) & = U_S(t) \otimes U_B(t)
\notag
\\
& \equiv \sum_{n} e^{i\phi_n(t)}\ket{n(t)} \bra{n(0)} \otimes \prod_{i = 1}^N e^{-iH_{B_i}t},
\end{align}
Here, $\ket{n(t)}$ is the instantaneous eigenstate of the bare Hamiltonian $H_0(t)$, which satisfies
\begin{align}
H_0(t) \ket{n(t)}
= E_n(t)\ket{n(t)}.
\end{align} 
The phase factor $\phi_n(t)$ can be decomposed into two parts, dynamic phase $\eta_n(t)$ and adiabatic phase $\xi_n(t)$, whose definitions are as follows \cite{PhysRevA.104.0322012021, griffiths2018introduction},
\begin{align}
\eta_n(t) & = - \int_{0}^{t} E_n(\tau) d\tau,
\\
\xi_n(t) & =  i \int_{0}^{t} \bra{n(\tau)} \frac{d}{d\tau} \ket{n(\tau)} d\tau 
\notag
\\
& = i \int_{0}^{t} \braket{n(\tau)|\dot{n}(\tau)} d\tau .
\end{align}
For simplicity, we have set $\hbar = 1$ throughout the whole paper.

By virtue of the transformation operator given above, the density operator $\rho(t)$ of the total composite system and its corresponding operator $\tilde{\rho}(t)$ in the adiabatic frame are related according to $\tilde{\rho}(t) = U^{\dagger}(t) \rho(t) U(t)$. The Liouville equation of $\tilde{\rho}(t)$ then reads
\begin{align}
\frac{d}{dt} \tilde{\rho}(t) = -i [\tilde{H}(t), \tilde{\rho}(t)],
\label{eq:Liouville}
\end{align}
Here, $\tilde{H}(t)$ is the transformed total Hamiltonian 
{which can be divided into three parts,}
\begin{align}
\tilde{H}(t) = \tilde{H}_0(t) + \lambda \tilde{H}_I(t) + \zeta \sum_{i=1}^{N}\tilde{V}_i(t),
\end{align}
where $\tilde{H}_0(t)$ is the transformed bare Hamiltonian of the system given as
\begin{align}
\tilde{H}_0(t) = \sum_{n \neq m} \sum_{m}\alpha_{nm}(t) \ket{n(0)}\bra{m(0)}
\end{align} 
with $\alpha_{nm}(t)$ being
\begin{align}
\alpha_{nm}(t) & = -ie^{-i\left[ \phi_n(t) - \phi_m(t)\right] } \braket{n(t)|\dot{m}(t)}
\notag
\\
& \equiv -ie^{-i \phi_{nm}(t)} \braket{n(t)|\dot{m}(t)}.
\end{align}
$\tilde{H}_I$ and  $\tilde{V}$ are called transformed intrasubsystem interaction and transformed subsystem-bath coupling, whose definitions are
\begin{align}
\tilde{H}_I(t) = U^{\dagger}(t) H_I U(t)
\end{align}
and
\begin{align}
\tilde{V}_i(t) = U^{\dagger}(t) V_i U(t),
\label{eq:tildeV}
\end{align}
respectively. {Note that in the most general case, the form of the system-bath coupling $V_i$ can be expressed as \cite{HPBFP}
\begin{align}
V_i & = \sum_{\mu} A_{i\mu} \otimes  B_{i\mu} 
= \sum_{\mu} A_{i\mu}^{\dagger} \otimes  B_{i\mu}^{\dagger} .
\label{eq:V}
\end{align}
Therefore,  the transformed subsystem-bath coupling $\tilde{V}_i(t)$ takes the form
\begin{align}
\tilde{V}_i(t)  = & \sum_{\mu}  \sum_{n,m} e^{-i\phi_{nm}(t)} \mathcal{A}_{i\mu}(n,m,t)  \otimes \tilde{B}_{i\mu}(t)
\notag
\\
= &\sum_{\mu}  \sum_{n,m} e^{i\phi_{nm}(t)} \mathcal{A}^{\dagger}_{i\mu}(m,n,t)  \otimes \tilde{B}_{i\mu}^{\dagger}(t).
\label{eq:Vi}
\end{align}
Here, we have introduced the notations
\begin{align}
\mathcal{A}_{i\mu}(n,m,t) = & \left[ \mathcal{A}^{\dagger}_{i\mu}(m,n,t) \right]^{\dagger}
\notag
\\
\equiv &\bra{n(t)} A_{i\mu} \ket{m(t)} \ket{n(0)} \bra{m(0)}
\label{eq:aadagger}
\end{align}
and
\begin{align}
\tilde{B}_{i\mu}(t) \equiv e^{iH_{B_i}t}B_{i\mu} e^{-iH_{B_i}t}
\label{eq:tildeB}
\end{align}
for brevity.

Before making any approximations, we firstly apply a special iterative technique proposed in Ref.~\cite{RevModPhys.94.045006} to rewrite Eq.~(\ref{eq:Liouville}). This powerful technique plays an essential role in the derivation (see Appendix \ref{ap:ite} for detail). After taking the partial trace over the baths on both sides, the master equation of the system takes the following equivalent form,
\begin{align}
\frac{d}{dt} \tilde{\rho}_S(t) = & -i [\tilde{H}_0(t) + \lambda \tilde{H}_I(t), \tilde{\rho}_S(t)] -i\zeta \sum_{i=1}^{N} \Tr_B [\tilde{V}_i(t), \tilde{\rho}(0)] 
\notag
\\
& - \zeta\sum_{i=1}^{N} \int_{0}^{t}  ds \, \Tr_B [\tilde{V}_i(t),	[\tilde{H}(s), \tilde{\rho}(s)]].
\label{eq:12}
\end{align}
Here, $\tilde{\rho}_S(t)  = \Tr_B\left( \tilde{\rho}(t) \right) $ denotes the reduced density operator for the system.

We then perform the Born-Markov approximation to simplify the master equation. The Born approximation \cite{HMW, KJ, HPBFP}, states that the interaction between the subsystems and baths is weak enough, thus the state of the total composite system is described by a tensor product state, $\tilde{\rho}(t) \approx \tilde{\rho}_S(t) \otimes \prod_{i} \rho_{B_i}$. The Markov approximation \cite{HMW, KJ, HPBFP} assumes that the impact of system's past state on its future evolution can be neglected. Therefore, it is justified to replace $\tilde{\rho}_S(s)$ by $\tilde{\rho}_S(t)$ and change the lower limit of the integral in Eq.~(\ref{eq:12}) to be negative infinity \cite{HPBFP}. In addition, we also use the standard assumption $\Tr_B\left( B_{i\mu} \rho_{B_i}\right) = 0 $ \cite{RevModPhys.94.045006, HPBFP, PhysRevE.107.014108} to eliminate the cross terms in the master equation, such as $\Tr_B\left( \tilde{H}_0 \tilde{V}_i \rho_{B_i}\right) $, $\Tr_B\left( \tilde{H}_I \tilde{V}_i \rho_{B_i}\right) $, and $\Tr_B\left( \tilde{V}_i \tilde{V}_j \rho_{B_i}\right) (i \neq j)$. After all these procedures, we obtain the following Redfield-like master equation,
\begin{align}
\frac{d}{dt} \tilde{\rho}_S(t) =  -i [\tilde{H}_0(t) + \lambda\tilde{H}_I(t), \tilde{\rho}_S(t)] 
+ \zeta^2 \sum_{i = 1}^{N}\mathcal{L}_i[\tilde{\rho}_S(t)],
\label{eq:me}
\end{align} 
whose dissipation term $\mathcal{L}_i[\tilde{\rho}_S(t)]$ is given as
\begin{widetext}
\begin{align}
&\mathcal{L}_i[\tilde{\rho}_S(t)] 
\notag
\\
\approx & -\int_{0}^{+\infty} ds \, \Tr_B \left[ \tilde{V}_i(t),  \left[ 
\tilde{V}_i(t - s), \tilde{\rho}_S(t) \otimes \rho_{B_i} \right] \right]
\notag
\\
= & \sum_{\mu,n,m}  \sum_{\mu',n',m'} \int_{0}^{+\infty} ds  e^{i\left[ \phi_{n'm'}(t) - \phi_{nm}(t-s)\right] } 
C_{i\mu' \mu}(s) 
\left[ \mathcal{A}_{i\mu}(n,m,t-s) \tilde{\rho}_S(t)\mathcal{A}^{\dagger}_{i\mu'}(m',n',t)
- \mathcal{A}^{\dagger}_{i\mu'}(m',n',t)\mathcal{A}_{i\mu}(n,m,t-s) \tilde{\rho}_S(t)\right]  
 +  \text{\rm H.c.}.
\label{eq:markov}
\end{align}
\end{widetext}
Here, 
\begin{align}
C_{i\mu' \mu}(s)  \equiv  \Tr_B\left[\tilde{B}^{\dagger}_{i\mu'}(s) \tilde{B}_{i\mu}(0) \rho_{B_i}\right]
\label{eq:cf}
\end{align}
stands for the correlation function \cite{HMW, KJ, HPBFP} of bath $i$ and $\rm{H.c.}$ represents the Hermitian conjugate terms.

However, in contrast with the standard Redfield equation \cite{RevModPhys.94.045006, HMW, KJ, HPBFP}, the phase factor $\phi_{nm}(t-s)$ in $\mathcal{L}_i[\tilde{\rho}_S(t)] $ is typically not a straightforward linear function of time. Moreover, the jump operators like $\mathcal{A}_{i\mu}(n,m,t -s)$ are also complex time-dependent functions. Hence, the additional approximation is required to further simplify Eq.~(\ref{eq:me}) to make it applicable. Note that in the theoretical discussion, the correlation function $C_{i\mu' \mu}(s)$ is regarded as a nonzero function within the relaxation time $\tau_B$ while it decays to zero dramatically when $s \gtrsim \tau_B$ \cite{HPBFP}. That is, the integrand contributes to Eq.~(\ref{eq:markov}) mainly in the region $0 \lesssim s \lesssim \tau_B$. For a Markovian heat bath, $\tau_B$ is commonly believed to be short enough so that the phase factor is almost unchanged in this time scale. In addition, the bare Hamiltonian $H_0(t)$ also changes little over $\tau_B$ as has previously been assumed.} Therefore, it is justified to expand $\phi_{nm}(t-s)$ and $\mathcal{A}_{i\mu}(n,m,t-s)$ in the vicinity of $t$, retaining only terms up to the first order in $s$ \cite{PhysRevA.98.052129},
\begin{align}
& \phi_{nm}(t-s) \approx \phi_{nm}(t) -s \dot{\phi}_{nm}(t),
\label{eq:phi}
\\
& \mathcal{A}_{i\mu}(n,m,t-s) \approx \mathcal{A}_{i\mu}(n,m,t) - s \dot{\mathcal{A}}_{i\mu}(n,m,t).
\label{eq:a}
\end{align}
Here, we have introduced the notations $\dot{\phi}_{nm} \equiv d {\phi}_{nm}/dt$ and $\dot{\mathcal{A}}_{i\mu}\equiv d{\mathcal{A}}_{i\mu} /dt $ for simplicity.
{The whole dissipation term $\mathcal{L}_i[\tilde{\rho}_S(t)]$ then can be split into the zeroth-order terms $\mathcal{L}^{(0)}_i[\tilde{\rho}_S(t)]$ and the  corresponding first-order correction terms $\mathcal{L}^{(1)}_i[\tilde{\rho}_S(t)]$,
\begin{align}
\mathcal{L}_i[\tilde{\rho}_S(t)] = \mathcal{L}^{(0)}_i[\tilde{\rho}_S(t)] + \mathcal{L}^{(1)}_i[\tilde{\rho}_S(t)].
\label{eq:li}
\end{align}
The specific expressions for $\mathcal{L}^{(0)}_i[\tilde{\rho}_S(t)]$ and  $\mathcal{L}^{(1)}_i[\tilde{\rho}_S(t)]$ are given by Eqs.~(\ref{eq:l0}) and (\ref{eq:l1}) in Appendix \ref{ap:TDLME}, respectively. The zeroth-order dissipation rate $\Gamma^{(0)}_{i\mu'\mu} $ in $\mathcal{L}^{(0)}_i[\tilde{\rho}_S(t)]$ and the first-order dissipation rates $\Gamma^{(1)}_{i\mu'\mu}$ in $\mathcal{L}^{(1)}_i[\tilde{\rho}_S(t)]$ are defined as,}
\begin{align}
\Gamma^{(0)}_{i\mu'\mu} (\dot{\phi}_{nm}(t)) & \equiv \int_{0}^{+\infty} ds \,  e^{i s\dot{\phi}_{nm}(t)}
C_{i\mu' \mu}(s) ,
\label{eq:gamma0}
\\
\Gamma^{(1)}_{i\mu'\mu} (\dot{\phi}_{nm}(t)) & \equiv -\int_{0}^{+\infty} ds \,  e^{i s\dot{\phi}_{nm}(t)}
C_{i\mu' \mu}(s) s .
\label{eq:gamma1}
\end{align}

Until now, we indeed have obtained the most general form of the TDLME. For convenience in the practical application, we introduce the rotating wave approximation \cite{HMW, HPBFP} to further  simplify $\mathcal{L}_i[\tilde{\rho}_S(t)] $. {In this secular approximation, the terms for which $\phi_{n'm'}(t) \neq \phi_{nm}(t)$ are assumed to oscillate rapidly so that they can be neglected. However,} the simplified form of $\mathcal{L}_i[\tilde{\rho}_S(t)] $ under this assumption is still not universal. It rather depends on the energy spectrum of the system {and the process that the system has undergone. For simplicity, here we consider the simplest case where both the phase factors $\phi_{n}(t)$ and $\phi_{nm}(t)$ are non-degenerate} \footnote{The phase factor $\phi_{n}(t)$ is non-degenerate means that we have $n = m$ when $\phi_{nm}(t) = 0$, while $\phi_{nm}(t)$ is non-degenerate implies that we have $n = n'$ and $m = m'$ under the conditions of $\phi_{nm}(t) \neq 0$ and $\phi_{nm}(t) = \phi_{n'm'}(t)$. See the toy model discussed in Sec.~\ref{sec:example} as an example.}. In this situation, only the following two categories of terms are left: the terms with $n = m$, $n' = m'$ and the terms with $n = n'$ and $m = m'$. {By performing the secular approximation, the first-order dissipation term $\mathcal{L}^{(1)}_i[\tilde{\rho}_S(t)]$ can be reduced into a simpler form
[see Eq.~(\ref{eq:li1}) in Appendix \ref{ap:TDLME} for detail].
Regarding the zeroth-order dissipation term  $\mathcal{L}^{(0)}_i[\tilde{\rho}_S(t)]$, it is more customary to write it into the compact Lindblad-Gorini-Kossakowski-Sudarshan (LGKS) form \cite{HMW, KJ, HPBFP, lindblad1976generators, gorini1976completely, carmichael2013statistical} in the literature.} After introducing the parameters
\begin{align}
&\gamma^{(0)}_{i\mu'\mu} (\dot{\phi}_{nm}) \equiv \Gamma^{(0)}_{i\mu'\mu} + \left( \Gamma^{(0)}_{i\mu\mu'}\right) ^*,
\label{eq:gammaj}
\\
& S^{(0)}_{i\mu'\mu} (\dot{\phi}_{nm})\equiv \frac{1}{2i}\left[ \Gamma^{(0)}_{i\mu'\mu} - \left( \Gamma^{(0)}_{i\mu \mu'}\right) ^*\right], 
\label{eq:sj}
\end{align}
{$\mathcal{L}^{(0)}_i[\tilde{\rho}_S(t)]$ takes the following equivalent LGKS form,
\begin{align}
\mathcal{L}^{(0)}_i[\tilde{\rho}_S(t)] = & -i[H^{(0)}_{i,LS}, \tilde{\rho}_S(t)] + \mathcal{D}_i^{(0)}[\tilde{\rho}_S(t)].
\label{eq:final_li}
\end{align}
Here, $H^{(0)}_{i,LS}$ is known as the Lamb shift Hamiltonian and $\mathcal{D}_i^{(0)}[\tilde{\rho}_S(t)]$ represents the LGKS form dissipator \cite{HPBFP}. The concrete expressions of them are presented in Eqs.~(\ref{eq:lamb}) and (\ref{eq:d0}) in Appendix \ref{ap:TDLME}.}

Combining Eqs.~(\ref{eq:me}), (\ref{eq:li}), and (\ref{eq:final_li}), we finally derive the following TDLME in the adiabatic frame of reference,
\begin{align}
\frac{d}{dt} \tilde{\rho}_S = &  -i \left[ \tilde{H}_0(t) + \lambda\tilde{H}_I(t)+\zeta^2\sum_{i = 1}^{N}  H^{(0)}_{i,LS} , \tilde{\rho}_S(t)\right]  
\notag
\\
&+ \zeta^2 \sum_{i = 1}^{N} \left( \mathcal{D}_i^{(0)}[\tilde{\rho}_S(t)] + \mathcal{L}_i^{(1)}[\tilde{\rho}_S(t)]\right). 
\label{eq:TDLME}
\end{align}
{This is the first main result of our paper. In this equation, the intrasubsystem coupling $\lambda\tilde{H}_I$ only makes contribution to the unitary evolution of the system. In addition, the jump operators in each monomial of $\mathcal{D}_i^{(0)}[\tilde{\rho}_S(t)]$ and $\mathcal{L}_i^{(1)}[\tilde{\rho}_S(t)]$, such as $\mathcal{A}_{i\mu}(n,n,t)$ and $\mathcal{A}_{i\mu}^{\dagger}(n',n',t)$, act on the Hilbert space of the same subsystem. As a consequence, the interplay effect of two different
local heat baths is absent and Eq.~(\ref{eq:TDLME}) is truly a kind of local master equation.} By using the relation $\rho_S(t)= \Tr_B\left[ U(t) \tilde{\rho}(t) U^{\dagger}(t)\right] = U_S(t) \tilde{\rho}_S(t) U^{\dagger}_S(t) $, it is straightforward to obtain the TDLME in the Schr\"{o}dinger picture.

Note that when $H_0$ is time-independent, neglecting the Lamb shift contribution and then transforming Eq.~(\ref{eq:TDLME}) back into the original Schr\"{o}dinger picture lead to the standard LGKS form of the LME \cite{Levy_2014} (see Appendix~\ref{ap:recovery} for detail; see also the toy model discussed in Sec.~\ref{sec:example} as a specific example),
\begin{align}
\frac{d}{dt}\rho_S =  -i \left[  H_0 + \lambda  {H}_I, \rho_S(t) \right]  
+ \zeta^2 \sum_{i = 1}^{N}  \mathcal{D}_i^{(0)}[\rho_S(t) ] . 
\end{align}
Thus, Eq.~(\ref{eq:TDLME}) can be regarded as a generalization of the LME.

{\subsection{TDLME for bosonic heat baths} \label{sec:bosonic}
}
In the preceding subsection, we have derived the most general form of the TDLME. To exactly determine the expressions of the parameters $\Gamma^{(0)}_{i\mu'\mu}$ and $\Gamma^{(1)}_{i\mu'\mu}$ in the master equation, the nature of the baths as well as the structure of the system-bath interaction should be specified. In this section, we consider one of the most frequently adopted models in theoretical discussion, where the heat bath is composed of a multimode bosonic field (such as the electromagnetic field), and the subsystem couples to the bath via the dipole interaction \cite{HMW, HPBFP, PhysRevA.104.0322012021}. To be concrete, the bath's Hamiltonian $H_{B_i}$ and the subsystem-bath interaction $ V_i$ take the form
\begin{align}
H_{B_i} = \sum_{k_i} \omega_{k_{i}} b^{\dagger}_{k_{i}} b_{k_{i}}
\end{align}
and
\begin{align}
V_i & =  A_{i} \otimes B_i 
\notag
\\
& = A_{i} \otimes \sum_{k_{i}} g_{k_{i}} \left( b^{\dagger}_{k_{i}} + b_{k_{i}} \right),
\end{align}
respectively. Here, the label $k_{i}$ denotes the mode of the field for bath $i$. In addition, $\omega_{k_{i}}$ and $b_{k_{i}}$ are frequency and {bosonic} annihilation operator of the field mode, and $g_{k_{i}}$ is the coupling strength between the subsystem and the local heat bath.

We now start to evaluate the correlation function of the thermal bath. Without loss of the generality, we assume the operators $A_i$ and $B_i$ are Hermitian. In this case, $g_{k_{i}}$ is a real number. By applying the transformation given by Eq.~(\ref{eq:tildeB}), we obtain 
\begin{align}
\tilde{B}_i(t) = \sum_{k_{i}} g_{k_{i}}\left( b_{k_{i}}^{\dagger} e^{i\omega_{k_{i}} t} + b_{k_{i}} e^{-i\omega_{k_{i}} t} \right).
\label{eq:tildeBi}
\end{align}
Inserting it into Eq.~(\ref{eq:cf}) and after some rearrangement, the bath correlation function $C_i(s)$ takes the following form (see Appendix~\ref{ap:correlation} for detail), 
\begin{align}
&C_i(s) 
\notag
\\
= & \int_{0}^{+\infty}d \omega \,  \rho_i(\omega)g_i^2(\omega)
\left[ \coth\left(\frac{\beta_i \omega}{2} \right)\cos\left(\omega s \right) -i\sin\left(\omega s \right) \right] .
\end{align}
Here, $\rho_i(\omega)$ is the mode density of the field and $g_i(\omega)$ is the strength of the coupling under the continuous frequency limitation. Making use of the spectral density defined as
\begin{align}
\mathcal{J}_i(\omega) \equiv \rho_i(\omega)g_i^2(\omega),
\end{align} 
the bath correlation function can be rewritten into the standard form \cite{HPBFP, PhysRevA.104.0322012021} ,
\begin{align}
C_i(s) = \int_{0}^{+\infty}d \omega \,  \mathcal{J}_i(\omega)
\left[ \coth\left(\frac{\beta_i \omega}{2} \right)\cos\left(\omega s \right) -i\sin\left(\omega s \right) \right].
\label{eq:cis}
\end{align}
The combination of Eqs.~(\ref{eq:gamma0}), (\ref{eq:gammaj}), (\ref{eq:sj}), and (\ref{eq:cis}) immediately yields

\begin{align}
\gamma^{(0)}_{i}(\dot{\phi}) = & 2 \int_{0}^{+\infty} ds\int_{0}^{+\infty}d \omega \, \mathcal{J}_i(\omega) \bigg[ \sin\left(\omega s \right) \sin\left(\dot{\phi} s \right) 
\notag
\\
& + \coth\left(\frac{\beta_i \omega}{2} \right)\cos\left(\omega s \right) \cos\left(\dot{\phi} s \right)  \bigg] ,
\label{eq:gamma0i}
\\
S^{(0)}_{i} (\dot{\phi}) = & \int_{0}^{+\infty} ds\int_{0}^{+\infty}d \omega \,  \mathcal{J}_i(\omega) \bigg[ - \sin\left(\omega s \right) \cos\left(\dot{\phi} s \right)
\notag
\\
&  +  \coth\left(\frac{\beta_i \omega}{2} \right)\cos\left(\omega s \right) \sin\left(\dot{\phi} s \right)  \bigg] .
\label{eq:s0i}
\end{align}
Here, we have used the notation $\dot{\phi} = \dot{\phi}_{nm}(t)$ for the sake of brevity. {Given the spectral density $\mathcal{J}_i(\omega)$, these two integrals can be worked out analytically or numerically.} In the theoretical discussion, $\mathcal{J}_i(\omega)$ is {usually phenomenologically modeled as a power function of the frequency for small values of $\omega$, 
\begin{align}
\mathcal{J}_i(\omega)\propto \omega^{a}, (\omega \to 0).
\end{align}
Here, $0< a <1$, $a = 1$, and $a >1$ correspond to the sub-Ohmic, Ohmic, and super-Ohmic spectral
density, respectively \cite{RevModPhys.59.1}. In addition, the truncation functions, such as Lorentz-Drude and exponential cutoff functions \cite{HPBFP, PhysRevA.104.0322012021, RevModPhys.59.1}, are necessary to ensure that Eq.~(\ref{eq:cis}) is convergent. As a simple yet typical example,} here we take $\mathcal{J}_i(\omega)$ to be an Ohmic spectral density with a Lorentz-Drude cutoff function \cite{HPBFP},
\begin{align}
\mathcal{J}_i(\omega) = \frac{2\kappa_i}{\pi} \omega \frac{\Omega_i^2}{\Omega_i^2 + \omega^2},
\label{eq:spectral}
\end{align}
where $\kappa_i$ is a frequency-independent constant and $\Omega_i$ is the cutoff frequency of the spectral density. Furthermore, we assume that the temperature of the local bath is high enough so that $k_B T_i \gtrsim \Omega_i$ is satisfied. In this case, $\gamma^{(0)}_{i}$ and $S^{(0)}_{i}$ {can be analytically obtained with the help of the Matsubara expansion \cite{HPBFP}} (see Appendix~\ref{ap:parameter} for detail),
\begin{align}
&\gamma^{(0)}_{i}(\dot{\phi}) =  {\pi} \mathcal{J}_i(\dot{\phi}) \left( \coth \frac{\beta_i \dot{\phi}}{2} + 1\right) ,
\label{eq:1dissipation}
\\
& S^{(0)}_{i}(\dot{\phi}) \approx \frac{\kappa_i \Omega_i}{\dot{\phi}^2 + \Omega_i^2}\left( 2k_BT_i\dot{\phi} - \Omega_i^2\right).
\end{align}
{Note that
\begin{align}
\Gamma^{(1)}_i(\dot{\phi})  = i\frac{d}{d \dot{\phi}} \Gamma^{(0)}_{i}(\dot{\phi}),
\end{align}
the first-order dissipation rate $\Gamma^{(1)}_i(\dot{\phi})$ then takes the form, 
\begin{align}
\Gamma^{(1)}_i(\dot{\phi})  
\approx & -\frac{2\kappa_i \Omega_i}{(\dot{\phi}^2 + \Omega_i^2)^2}\left[  k_BT_i(\Omega_i^2 - \dot{\phi}^2) + \Omega_i^2 \dot{\phi} \right] 
\notag
\\
& + \frac{i\pi}{2} \left[  \mathcal{J}'_i(\dot{\phi})\left( \coth \frac{\beta_i \dot{\phi}}{2} + 1\right) - \frac{\beta_i \mathcal{J}_i(\dot{\phi})}{2\sinh ^2 \frac{\beta_i \dot{\phi}}{2}}  \right] .
\end{align}
Thus, by specifying a particular spectral density, we have analytically determined the form of the zeroth- and first-order dissipation rates in the TDLME.}

\section{Thermodynamic inconsistencies for the LME and TDLME}
\label{sec:condition}

LME serves as an effective and widespread description for a time-independent boundary-driven system, whose subsystems interact with each other and weakly couple to their respective local heat bath \cite{PhysRevLett.105.130401, Levy_2014, PhysRevE.107.034128, PhysRevE.107.044102, PhysRevA.89.022128}. Nevertheless, it has been shown that the LME can lead to an apparent inconsistency with the second law of thermodynamics \cite{Levy_2014, vcapek2002zeroth, stockburger2017thermodynamic}. Despite great effects based on other alternative frameworks are made to reconcile the LME with the  thermodynamic law \cite{Levy_2014, Trushechkin_2016, barra2015thermodynamic, De_Chiara_2018, RevModPhys.93.035008, PhysRevResearch.3.013165, PhysRevE.107.014108}, the substantial reason behind this contradiction remains obscure. In the present section, we address this issue with theoretical analysis and establish the range of validity for the LME. 

As mentioned previously, the evolution of a system coupling to the local thermal reservoirs at its boundary can be characterized by the following LME \cite{De_Chiara_2018, PhysRevResearch.3.013165},
\begin{align}
\frac{d}{dt}\rho_S = &  -i \left[ \sum_{i} H_{s_i} + \lambda  {H}_I , \rho_S(t) \right]  + \zeta^2 \sum_{i}  \mathcal{D}_i [\rho_S(t) ] 
\label{eq:LME}
\end{align}
with dissipators
\begin{align}
\mathcal{D}_i [\rho_S(t)] 
=  \sum_{n_i,m_i} \gamma_{n_i,m_i}\left( L_{n_i,m_i} \rho_S L_{n_i,m_i}^{\dagger} - \frac{1}{2}\left\lbrace L_{n_i,m_i}^{\dagger}L_{n_i,m_i}, \rho_S  \right\rbrace \right) .
\end{align}
Here, $L_{n_i,m_i} = L_{m_i,n_i}^{\dagger}$ is the eigenoperator \cite{De_Chiara_2018} of $H_{s_i}$ which satisfies 
\begin{align}
\left[ H_{s_i}, L_{n_i,m_i} \right] = \left( \epsilon_{n_i} - \epsilon_{m_i} \right) L_{n_i,m_i}.
\end{align}
$\epsilon_{n_i}$ and $\epsilon_{m_i}$ are eigenvalues of $H_{s_i}$ \footnote{If we denote the eigenvector of $H_{s_i}$ corresponding to the eigenvalue $\epsilon_{n_i}$ by $\ket{n_i}$, then eigenoperator $L_{n_i,m_i}$ can be expressed as $L_{n_i,m_i} = \ket{n_i} \bra{m_i}$.}. The dissipation rates $\gamma_{n_i,m_i}$ and $\gamma_{m_i,n_i}$ obey the detailed balance condition,
\begin{align}
\frac{\gamma_{n_i,m_i}}{\gamma_{m_i,n_i}} = e^{-\beta_i
\left( \epsilon_{n_i} - \epsilon_{m_i} \right) },
\label{eq:dbc}
\end{align}
with $\beta_i \equiv 1/\left( k_B T_i\right) $ being the inverse temperature of bath $i$ and $k_B$ being Boltzmann constant. Conventionally, the Lamb shift term has been ignored for simplicity.

Let us start by temporarily considering the particular situation where the intrasubsystem interaction $\lambda H_I$ is absent. In this case, the subsystems are isolated with each other and they reach to their own thermal equilibrium state after a sufficient long time. Hence, the steady state of the system $\rho_{0,ss}$ can be denoted as
\begin{align}
\rho_{0,ss} = \otimes \prod_{i = 1}^N \frac{e^{-\beta_i H_{s_i}}}{Z_i},
\end{align}
where $Z_i = \Tr\left[  \exp\left( -\beta_i H_{s_i}\right) \right] $ is the partition function of subsystem $s_i$. The heat current flowing from bath $i$ to subsystem $s_i$ \cite{Levy_2014} at the steady state of course equals zero, i.e.,
\begin{align}
J_{0,i,ss} & \equiv \zeta^2\Tr\left( \sum_{k} H_{s_k} \mathcal{D}_i [\rho_{0,ss} ] \right)
\notag
\\
& = 0. 
\end{align}

We now go back to the general case by taking into account the impact of the  interaction between subsystems. Due to the presence of the intersubsystem coupling $\lambda H_I$, the steady state $\rho_{ss}(\lambda)$ of Eq.~(\ref{eq:LME}) now deviates from $\rho_{0,ss}$. We further suppose that $\rho_{ss}(\lambda)$ is a smoothly varying function of parameter $\lambda$ \footnote{This assumption does not necessarily hold true in certain specific cases, as discussed in detail in  Sec.~\ref{subsec:numerical}. In the present paper, however, we focus on the scenarios where this assumption is satisfied.}. Since the interaction $\lambda H_I$ is weak, the deviation is not so large and $\rho_{ss}(\lambda)$ can be approximated by the Taylor expansion up to the first order of $\lambda$,
\begin{align}
\rho_{ss}(\lambda) & \approx \rho_{ss}(0) + \lambda \rho^{(1)}
\notag
\\
& =  \rho_{0,ss} + \lambda \rho^{(1)},
\end{align}
where $\rho^{(1)}$ is the first-order correction term induced by the subsystem interaction. The heat current of bath $i$ \cite{Levy_2014} at the steady state becomes 
\begin{align}
J_{i,ss} \equiv & \zeta^2\Tr\left\lbrace \left(\sum_{k} H_{s_k} + \lambda H_I\right) \mathcal{D}_i [\rho_{ss}(\lambda) ] \right\rbrace 
\notag
\\
= & \zeta^2\Tr\left\lbrace \left(\sum_{k} H_{s_k} + \lambda H_I\right) \left( \mathcal{D}_i [\rho_{0,ss}] + \lambda  \mathcal{D}_i [\rho^{(1)}] \right)  \right\rbrace 
\notag
\\
= & \lambda\zeta^2 \Tr \left( \sum_{k} H_{s_k}\mathcal{D}_i [\rho^{(1)}] + H_I \mathcal{D}_i [\rho_{0,ss}]\right) + O(\lambda^2\zeta^2).
\label{eq:current} 
\end{align}
Here, we have used the linearity of superoperator $\mathcal{D}_i$.  Equation (\ref{eq:current}) shows that the leading order of the heat current $J_{i,ss}$ evaluated by the LME is {at most} $\lambda\zeta^2$. However, the LME itself is only kept up to the order of $\zeta^2$ and the higher order terms are completely neglected. This implies that the heat current $J_{i,ss}$ obtained from the LME {is typically not indicative of the actual}  steady-state heat current, and it is {therefore} not surprising that using this inaccurate heat current can lead to the violation of the second thermodynamic law. In other words, without carefully comparing the heat current $J_{i,ss}$ and its counterpart evaluated from the neglected dissipation terms, the LME inherently fails to precisely describe the thermodynamic properties near the steady state, even if the results obtained from this equation may coincidentally be consistent with the second thermodynamic law. The analysis made above focuses on the case in which each subsystem is coupled to its own heat bath. The situation where only a part of subsystem contacting with the local heat bath is discussed in Appendix~\ref{ap:parts}, and the conclusion is exactly the same as that drew here.

Although the LME in Eq.~(\ref{eq:LME}) {may lead to} the thermodynamic inconsistency {as the system nears steady state}, {it effectively} describes the evolution of the system {within its range of applicability, and the second law remains valid in this regime. According to the previous statement, the range in which the LME is valid corresponds to the condition that the  thermodynamic observables evaluated accordingly are at least of second-order small terms. Let us consider a state $\rho_S(t)$ in this scope, that is,} all the relevant thermodynamic quantities, like heat current $J_i(t)$ and entropy production rate $\dot{\Sigma}(t)$, are of the order of $\zeta^2$, 
\begin{align}
J_i(t) & \equiv   \zeta^2\Tr\left\lbrace \left(\sum_{k} H_{s_k} + \lambda H_I\right) \mathcal{D}_i [\rho_S(t) ] \right\rbrace \sim \zeta^2,
\label{eq:jit}
\\
\dot{\Sigma}(t) & \equiv \frac{dS}{dt} - \sum_i \beta_i J_i(t)  \sim \zeta^2.
\end{align}
Here, we have defined $S(t) \equiv -  \Tr\left[ \rho_S(t) \ln \rho_S(t) \right]  $ as the entropy of the system. It is convenient to divide the heat current $J_i(t)$ into two parts which have different orders of magnitude,
\begin{align}
J_i(t) & = \zeta^2\Tr\left( \sum_{k} H_{s_k}  \mathcal{D}_i [\rho_S(t) ] \right) +  \lambda\zeta^2\Tr\left(  H_{I}  \mathcal{D}_i [\rho_S(t) ] \right) 
\notag
\\
& \equiv J_i^s(t) + J_i^I(t) .
\end{align}
Here, $J_i^s$ and $J_i^I$ represent the contributions to the heat flow from the bare system Hamiltonian and the internal interaction, respectively. Since $J_i(t) \sim \zeta^2$ and $J_i^I(t) \sim \lambda\zeta^2$, the predominant contribution to the heat current $J_i(t)$ must come from the bare system Hamiltonian, namely $J_i(t) \approx  J_i^s(t) \sim \zeta^2$. Therefore,it is justified to approximate the entropy production rate $\dot{\Sigma}(t)$ by
\begin{align}
\dot{\Sigma}(t) & \approx \frac{dS}{dt} - \sum_i \beta_i J^s_i(t).  
\end{align}
According to the analysis in Refs.~\cite{RevModPhys.94.045006, PhysRevResearch.3.013165}, the entropy production rate which only involves the heat current associated with the bare system Hamiltonian is always non-negative,
{i.e.,} $\dot{\Sigma} \geqslant 0$, and the second law of thermodynamics holds in this case. {This provides us with a criterion to determine whether the LME follows the second thermodynamic law.

Now, let us make a brief summary of the conclusions we have drawn in this section. When the LME is applied within its applicable scope, such that the thermodynamic observables evaluated become second-order infinitesimals, it does not lead to the thermodynamic inconsistency. However, when we utilize the LME to characterize the steady-state properties, we typically find that the thermodynamic observables are higher-order terms relative to the small quantity presented in the equation. Consequently, it is unjustifiable to apply the LME in this context, even if the meticulous parameter tuning can circumvent the apparent violation of the second thermodynamic law.
}

The subtleties arise when we study the issue of the time-dependent boundary-driven system. In this case, even if we suppose that $ U_S\mathcal{A}_{i\mu}  U_S^{\dagger}$ is the instantaneous eigenoperator of $H_{s_i}(t)$, the operator $ U_S\dot{\mathcal{A}}_{i\mu} U_S^{\dagger}$ in the first-order dissipation term $ U_S\mathcal{L}^{(1)}_i U_S^{\dagger}$ does not have to be that way. To simplify the problem, here we consider the slowly driven case in which the characteristic time of the external driving is much smaller than the relaxation time of the system. Consequently, the term $\mathcal{L}^{(1)}_i$ can be regarded as the higher order perturbation compared with $\mathcal{D}^{(0)}_i$ term. Despite this, the detailed balance condition given by Eq.~(\ref{eq:dbc}) may not necessarily be satisfied during the evolution, due to the presence of the adiabatic phase $\xi_n$ [see Eq.~(\ref{eq:1dissipation}) as a typical example]. That is, generally speaking, the role of the heat bath in this case is not to thermalize the system to an equilibrium state, even though we ignore the first order correction terms $\mathcal{L}_i^{(1)}$. From this point of view, it makes little sense to discuss whether the second law is violated in this scenario. To make the investigation meaningful, we narrow our discussion down to the situation where we can find the parameters $\beta^{\rm eff}_i$ called effective temperature, such that the following equation,
\begin{align}
\frac{\gamma^{(0)}_{i}(\dot{\phi}_{n_i m_i})}{\gamma^{(0)}_{i}(\dot{\phi}_{m_i n_i})} = \exp{\left\lbrace -\beta^{\rm eff}_i
\left[  \epsilon_{n_i}(t) - \epsilon_{m_i}(t) \right] \right\rbrace  },
\label{eq:detail}
\end{align}
is satisfied for arbitrary $n_i$ and $m_i$ during the whole process concerned. Moreover, we further assume that $\beta^{\rm eff}_i$ is always equal to the temperature of the local heat bath, namely $\beta^{\rm eff}_i = \beta_i$. These two assumptions are simultaneously satisfied at the vanishing adiabatic phase $\xi_n$ [see Eq.~(\ref{eq:1dissipation}) in Sec.~\ref{sec:bosonic} as an example]. Under these premises, the TDLME derived in the present paper may also cause the disagreement with the second law when the system is close to the instantaneous steady state. The reason is that, for those {higher-order infinitesimals} of $\zeta^2$, only a portion of them are retained in Eq.~(\ref{eq:TDLME}), {while} the other parts, such as the terms proportional to $\lambda \zeta^2$, are {still} totally ignored as before. Thus, it may also give rise to the contradiction of the thermodynamic law when the system approaches the steady state. In analogy with the time-independent case, the TDLME works well {in its applicable scope}. In this situation, the heat current $J_i(t)$ and the entropy production rate $\dot{\Sigma}(t)$ are of the order of $\zeta^2$. This guarantees the chief contribution of the bare system Hamiltonian in the heat current. Therefore, the entropy production rate is always non-negative and the second law is obeyed.

\section{numerical illustration}
\label{sec:example}
{\subsection{Toy model}}

\begin{figure}[tb!]
	
\centering \includegraphics[scale=0.35]{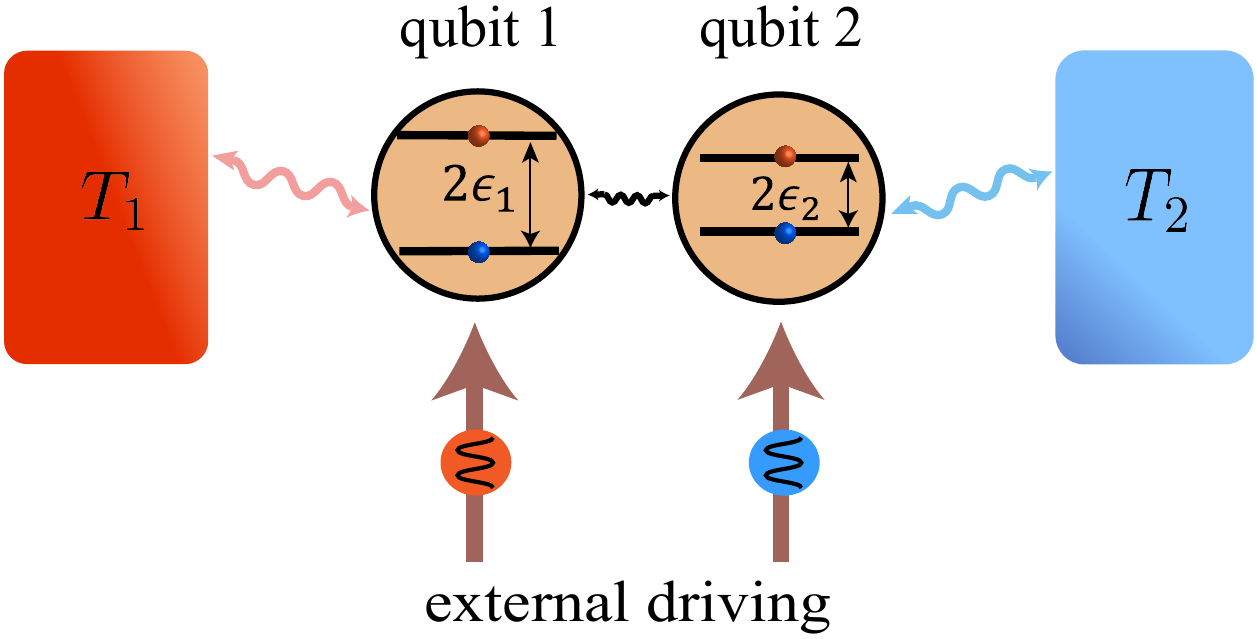}
	
\caption{Schematic diagram for the toy model. The system of interest consists of a pair of externally driven qubits. The two qubits interact with each other and the energy gaps for each qubit at the initial time are denoted by $2\epsilon_1$ and $2 \epsilon_2$, respectively. In addition, qubit $1$ is connected with the local hot heat bath while qubit $2$ is coupled to the local cold thermal reservoir.}	
\label{fig:toymodel}
\end{figure}

In this section, we confirm our results obtained previously by a toy model. As shown in Fig.~\ref{fig:toymodel}, the system concerned contains a pair of weakly coupled subsystems, qubit $1$ and qubit $2$, and both of them undergo the time-dependent external driving. The full Hamiltonian of the system reads
\begin{align}
H_S(t) = \sum_{i =1,2} \left[ {\epsilon_i} \sigma^z_i + f_i(t) \sigma^x_i\right] + \lambda \left(  \sigma_1^+ \sigma_2^- + \sigma_1^- \sigma_2^+ \right) .
\end{align}
Here, $\sigma_i^{x,y,z}$ represent Pauli matrix for qubit $i$, $\epsilon_i$ characters the energy gap for each bare Hamiltonian at the initial time, $f_i(t)$ and $\lambda$ denote the strengths of the external driving \footnote{Without loss of the generality, here we assume $f_i(0) = 0$.} and intrasubsystem interaction, respectively, and $\sigma_i^{\pm} \equiv \left( \sigma_i^x \pm i \sigma_i^y\right) /{2}$ are the raising and lowering operators for qubit $i$. In addition, each qubit connects to its own local {bosonic} heat bath via the following dipole interaction \cite{PhysRevA.104.0322012021}, 
\begin{align}
\zeta V_i  =   \zeta \sigma^x_{i} \otimes \sum_{k} g_{i,k} \left( b^{\dagger}_{i,k} + b_{i,k} \right).
\end{align}
Without loss of the generality, we have assumed $T_1 >T_2$.

For a two level system, it is convenient to introduce the basis $\ket{\uparrow}_i \equiv (1, 0)^T_i$ and $\ket{\downarrow}_i \equiv (0, 1)^T_i$ to rewrite the Pauli matrix as \cite{PhysRevA.104.0322012021}
\begin{align}
\sigma_i^x = \ket{\uparrow}_i\bra{\downarrow}_i + \ket{\downarrow}_i\bra{\uparrow}_i
\end{align}
and
\begin{align}
\sigma_i^z = \ket{\uparrow}_i\bra{\uparrow}_i - \ket{\downarrow}_i\bra{\downarrow}_i.
\end{align}  
The instantaneous eigenstates of each qubit, $\ket{e(t)}_i$ and $\ket{g(t)}_i$, in this representation take the form \cite{PhysRevA.104.0322012021},
\begin{align}
\ket{e(t)}_i = \cos \frac{\theta_i(t)}{2} \ket{\uparrow}_i + \sin \frac{\theta_i(t)}{2} \ket{\downarrow}_i
\end{align}
and 
\begin{align}
\ket{g(t)}_i = -\sin \frac{\theta_i(t)}{2} \ket{\uparrow}_i + \cos \frac{\theta_i(t)}{2} \ket{\downarrow}_i,
\end{align}
where we have introduced the parameters $\theta_i(t) \in \left(  -\pi/2, \pi/2 \right)  $ satisfying
\begin{align}
\tan \theta_i(t) = \frac{f_i(t)}{\epsilon_i}.
\end{align}
The corresponding eigenvalues are
\begin{align}
E_{e,i}(t) = -E_{g,i}(t) = \sqrt{\epsilon_i^2 + f_i^2(t)}.
\end{align}
Thanks to
\begin{align}
\sideset{_i}{_i}{\mathop{\braket{e(t)|\dot{e}(t)}}} = \sideset{_i}{_i}{\mathop{\braket{g(t)|\dot{g}(t)}}} =0,
\end{align}
all the adiabatic phases are zero and only dynamic phases are retained in the phase factor. Performing the adiabatic transformation defined in Sec.~\ref{sec:derivation} on the coupling $ \zeta V_i$ leads to \cite{PhysRevA.104.0322012021}
\begin{align}
 \zeta \tilde{V}_i(t) & =  \zeta U^{\dagger}(t) V_i U(t)
\notag
\\
& =   \zeta \left\lbrace \sin \theta_i {\sigma}_i^z + \cos \theta_i \left[  e^{-i\eta_{eg,i}(t)} {\sigma}_i^+ + {\rm H.c.} \right]  \right\rbrace \otimes \tilde{B}_i(t),
\label{eq:interaction}
\end{align}
where the difference of the dynamic phase $\eta_{eg,i}(t)$ is defined as
\begin{align}
\eta_{eg,i}(t) & \equiv \eta_{e,i}(t) - \eta_{g,i}(t)
\notag
\\
& =  -\int_{0}^{t} \left[ E_{e,i}(\tau) - E_{g,i}(\tau) \right] d\tau
\notag
\\
& =  - 2\int_{0}^{t} \sqrt{\epsilon_i^2 + f_i^2(\tau)} d\tau,
\end{align}
and the definition of $\tilde{B}_i(t)$ is given by Eq.~(\ref{eq:tildeBi}).

Combining Eqs.~(\ref{eq:TDLME}) and (\ref{eq:interaction}) and after some algebra, we finally obtain the following TDLME in the Schr\"{o}dinger picture,
\begin{align}
\frac{d\rho_S}{dt} = &  -i \left[  H_{S}(t) , \rho_S \right] + \zeta^2 \sum_{i = 1,2}\mathcal{L}_i[{\rho}_S(t)] 
\notag
\\
= &  -i \left[  H_{S}(t) , \rho_S \right]  + \zeta^2 \sum_{i = 1,2}  \bigg[ \gamma_i^z(t)\left( \hat{\sigma}_i^z \rho_S \hat{\sigma}_i^z - \rho_S \right) 
\notag
\\
&  +  \gamma_i^-(t) \left( \hat{\sigma}_i^- \rho_S \hat{\sigma}_i^+ - \frac{1}{2}\left\lbrace \hat{\sigma}_i^+ \hat{\sigma}_i^-, \rho_S \right\rbrace  \right)  
\notag
\\
&  + \gamma_i^+(t) \left( \hat{\sigma}_i^+ \rho_S \hat{\sigma}_i^- - \frac{1}{2}\left\lbrace \hat{\sigma}_i^- \hat{\sigma}_i^+, \rho_S \right\rbrace  \right)  \bigg] .
\label{eq:example}
\end{align}
Here, the operators $\hat{\sigma}_i^{\pm,z}$ are defined based on the instantaneous eigenstates; e.g., $\hat{\sigma}_i^{+} = \ket{e(t)}_i \bra{g(t)}_i$. The dissipation rates take the form,
\begin{align}
\gamma_i^z(t) & = \gamma_i^{(0)}(0)\sin^2 \theta_i + \gamma_i^{(1)}(0)\sin \theta_i \frac{d}{dt} \sin 
\theta_i,
\\
\gamma_i^-(t) & = \gamma_i^{(0)}(2E_{e,i})\cos^2 \theta_i + \gamma_i^{(1)}(2E_{e,i})\cos \theta_i  \frac{d}{dt} \cos \theta_i ,
\\
\gamma_i^+(t) &= \gamma_i^{(0)}(-2E_{e,i})\cos^2 \theta_i + \gamma_i^{(1)}(-2E_{e,i})\cos \theta_i  \frac{d }{dt}\cos \theta_i ,
\end{align}
with $\gamma_i^{(1)}(x)$ being
\begin{align}
\gamma_i^{(1)}(x) \equiv \Gamma^{(1)}_{i}(x) + \left( \Gamma^{(1)}_{i}(x)\right) ^*.
\end{align}
In addition, we have neglected the Lamb shift term as well as its corresponding modification part for simplicity. Note that when the Hamiltonian $H_S$ is time-independent, we immediately have $\gamma_i^z = 0$ and {$\hat{\sigma}_i^{\pm} = {\sigma}_i^{\pm}$}, and Eq.~(\ref{eq:example}) {reverts} to the conventional form of the LME \cite{Levy_2014}. Since the time derivative of the entropy takes the form \cite{PhysRevResearch.3.013165},
\begin{align}
\frac{dS}{dt} = - \zeta^2 \sum_{i = 1,2}\Tr\left\lbrace  \mathcal{L}_i[{\rho}_S(t)] \ln {\rho}_S(t)  \right\rbrace , 
\end{align}
the entropy production rate $\dot{\Sigma}(t)$ reads
\begin{align}
\dot{\Sigma}(t) & = \frac{dS}{dt} - \sum_i \beta_i J_i(t)
\notag
\\
& = -\zeta^2 \sum_{i = 1,2}\Tr\left\lbrace \left[ \ln {\rho}_S(t) + \beta_i H_S(t) \right] \mathcal{L}_i[{\rho}_S(t)]  \right\rbrace.  
\end{align}
{This physical quantity can be used to judge whether the second law of thermodynamics is violated.}

\begin{figure}[t!]
\centering  \includegraphics[scale=0.415]{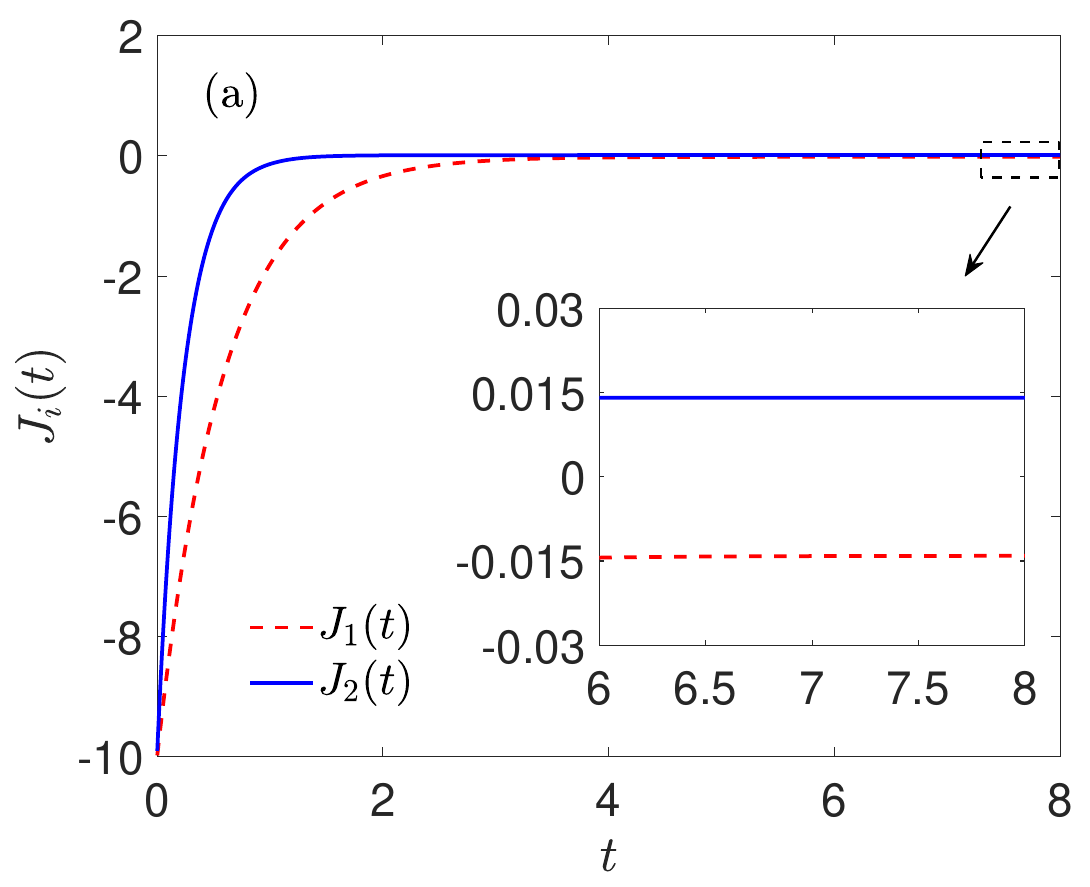}
	
\centering \includegraphics[scale=0.415]{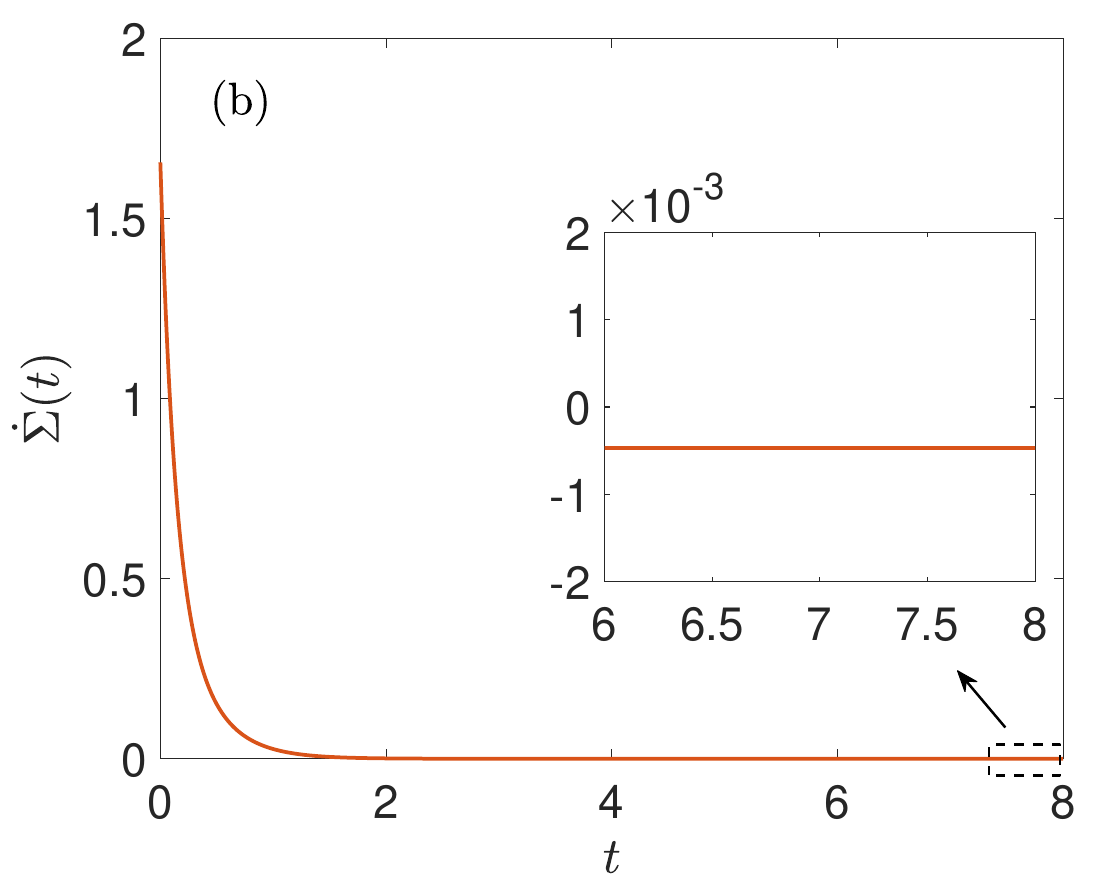}
\caption{Time evolution of (a) the heat currents $J_1(t)$ (red dashed curve), $J_2(t)$ (blue solid curve), and (b) the entropy production rate $\dot{\Sigma}(t)$ for the system without the external driving. Insets: The heat spontaneously flows from the hot bath to the cold bath ($J_1 < 0$ and $J_2 > 0$) at the steady state, and the entropy production rate becomes negative in this region, which implies the violation of the second law. Other parameters are set as: $k_B = 1$, $\kappa_1 = \kappa_2 = 10$, $\Omega_1 = \Omega_2 = 1$.}	
\label{fig:undriven}
\end{figure}

{\subsection{numerical simulation}
\label{subsec:numerical}
Let us focus our discussion on the most fundamental specific case where there is no external driving acting on the system. In this scenario, both analytical and numerical analyses can be conducted on the LME, which yields many meaningful and easily verifiable results. Note that the Hamiltonian of the system and the jump operator in the dissipation are quadratic and linear in the $\sigma_i^{\pm}$, respectively. The system satisfying this condition is typically referred to as Gaussian or noninteracting system in the literature \cite{RevModPhys.94.045006, De_Chiara_2018}. The dynamics of such a system is
entirely captured by the so-called covariance matrix, and the steady state solution of the Gaussian system can be analytically or numerically found by solving the Lyapunov equation \cite{RevModPhys.94.045006, Levy_2014, De_Chiara_2018}. The detailed analysis of the Gaussian system is given in Appendix \ref{ap:Lyapunov}. Here, we only present the final results. Figure \ref{fig:condition} shows the the dependency of steady-state entropy production rate $\dot{\Sigma}_{ss}$ on the heat bath temperature $T_i$ and the energy gap $\epsilon_i$. From this graph, we can clearly see that the LME gives rise to the thermodynamic contradiction when $\epsilon_1 / \epsilon_2 > T_1 / T_2$. This is exactly consistent with the results obtained in Ref.~\cite{Levy_2014}.

\begin{figure}[tb!]
\centering \includegraphics[scale=0.41]{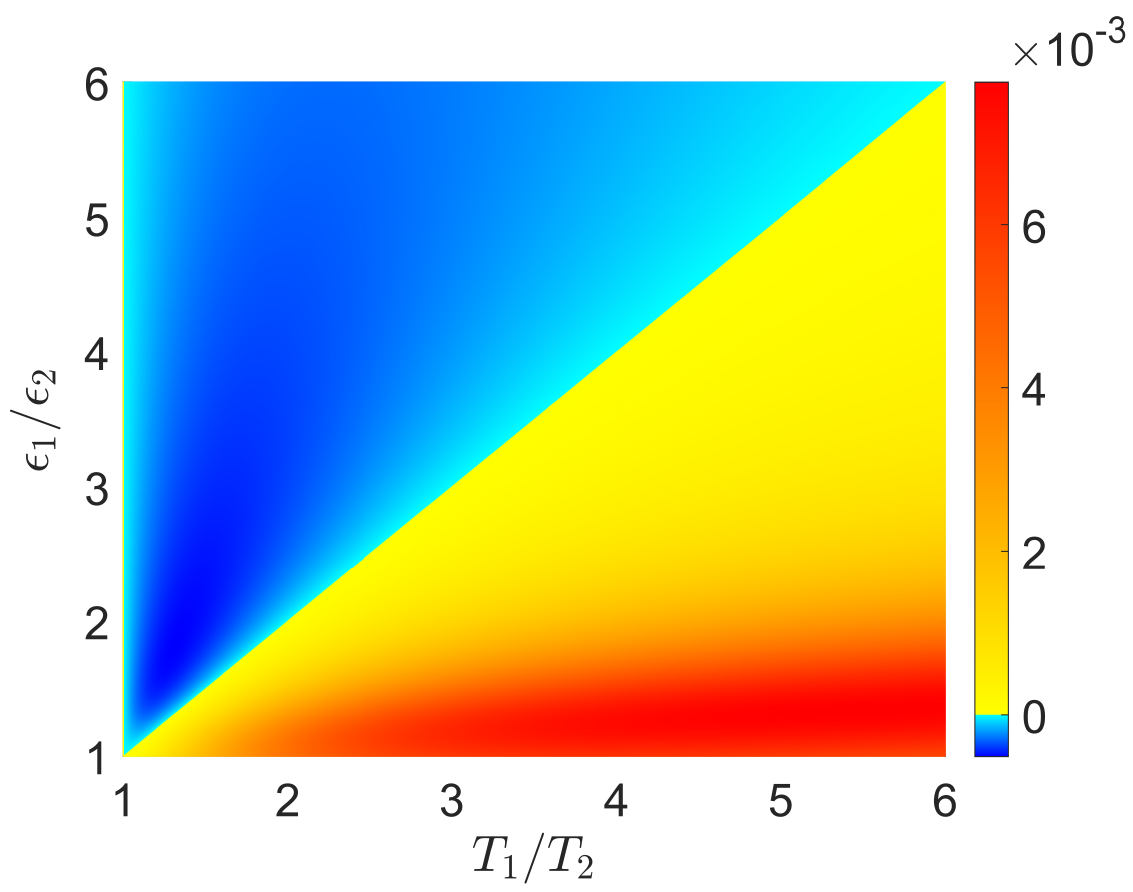}
	
\caption{Steady-state entropy production rate $\dot{\Sigma}_{ss}$ as a function of $T_1/T_2$ and $\epsilon_1 / \epsilon_2$. The blue area signifies a negative entropy production rate, which indicates the violation of the second law of thermodynamics. The borderline with a zero entropy production rate corresponds to $\epsilon_1/\epsilon_2 = T_1/T_2$. The parameters are set as: $k_B = 1$, $\kappa_1 = \kappa_2 = 10$, $\Omega_1 = \Omega_2 = 1$, $\lambda = \zeta^2 = 0.5$, $\epsilon_2 = 5$, $T_2 = 10$.}	
\label{fig:condition}
\end{figure}

In Sec.~\ref{sec:condition}, under the assumption of smooth variation of $\rho_{ss}(\lambda)$ with respect to $\lambda$, we have demonstrated that the steady-state heat current $J_{i,ss}$ evaluated by the LME are indeed higher-order infinitesimals of $\zeta^2$. However, it is crucial to emphasize that this conclusion may no longer hold true if the assumption of smoothness  is violated. In fact, in the system under consideration, both the smoothness of $\rho_{ss}(\lambda)$ and the magnitude of $J_{i,ss}$ are significantly affected by the relative order of magnitude among $\Delta \epsilon \equiv \epsilon_1 - \epsilon_2$, $\lambda$, and $\zeta^2$ \cite{Levy_2014, De_Chiara_2018}. As depicted in Fig.~\ref{fig:detuning}, under the conditions that $\lambda \sim \zeta^2$ and $\Delta \epsilon \ll \zeta^2$, $J_{1,ss}$ and $\zeta^2$ are of the same order of magnitude. This implies that $\rho_{ss}(\lambda)$ in this case is a sharply varying function of $\lambda$ \cite{De_Chiara_2018}. Additionally, the LME does not introduce the thermodynamic inconsistency in this scenario, as the system indeed absorbs the heat from the hot heat bath (i.e., $J_{1,ss} > 0$) at the steady state. This numerical result is fully in accordance with our
criterion that the LME is consistent with the second law of thermodynamics, when applied within its appropriate scope. As $\Delta \epsilon$ increases, $J_{1,ss}$ decreases dramatically. Notably, {when $\Delta \epsilon \gtrsim 0.9$, $J_{1,ss}$ falls below 0.01, which is a higher-order small term compared to $\zeta^2$. This suggests that $\rho_{ss}(\lambda)$ varies smoothly with $\lambda$ in this region. Furthermore, as $\Delta \epsilon$ continues to increase (specifically, when $\Delta \epsilon > 5$), $J_{1,ss}$ turns negative, indicating} a violation of the second law of thermodynamics. In the following discussion, we will concentrate on the scenario where { $\rho_{ss}(\lambda)$ is a smooth varying function of $\lambda$.}

\begin{figure}[tb!]
\centering
\vspace{3.5mm} \includegraphics[scale=0.4]{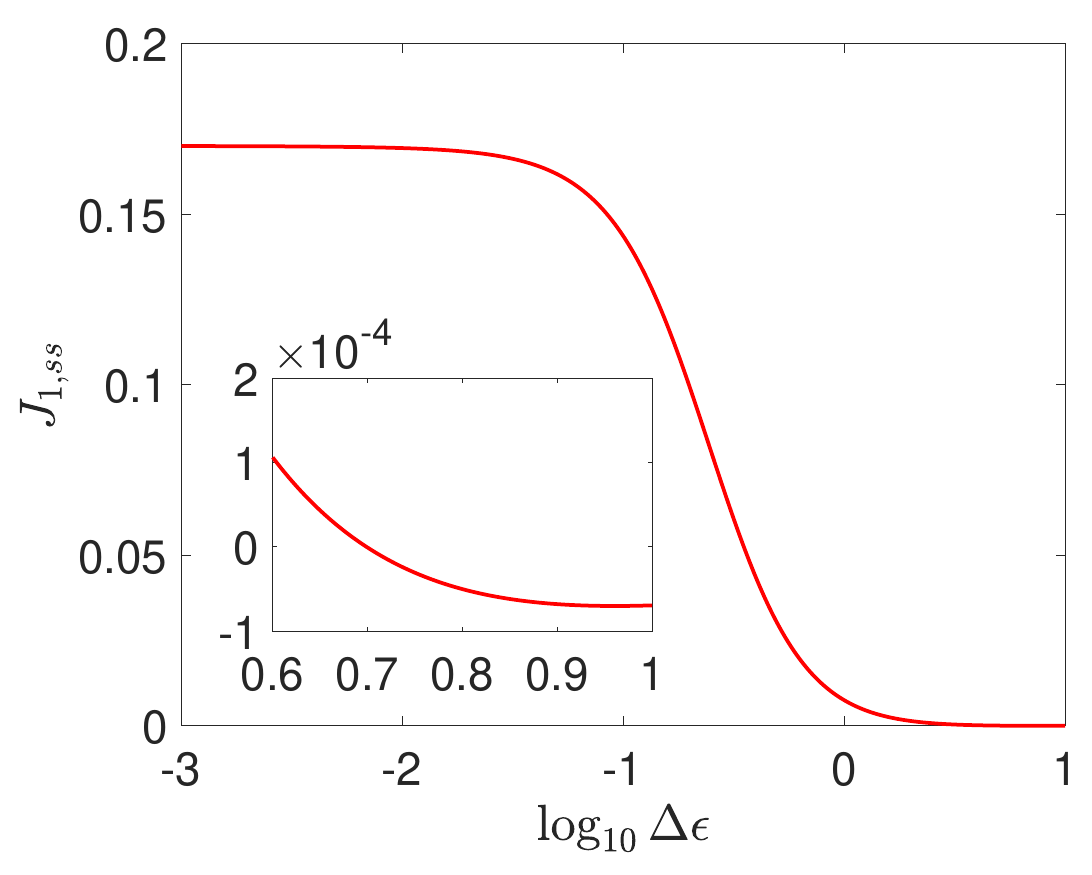}

\caption{Steady-state heat current $J_{1,ss}$ as a function of the energy gap difference $\Delta \epsilon$ between two qubits. Given the conditions where $\lambda \sim \zeta^2$ and $\Delta \epsilon \ll \zeta^2$, the order of magnitude of $J_{1,ss}$ is comparable to that of $\zeta^2$. In addition, the LME is consistent with the second thermodynamic law in this particular scenario. Inset: The heat current $J_{1,ss}$ turns negative and becomes a higher-order infinitesimal of  $\zeta^2$ when $\Delta \epsilon \gg \zeta^2$. The parameters are chosen as: $k_B = 1$, $\kappa_1 = \kappa_2 = 10$, $\Omega_1 = \Omega_2 = 1$, $\lambda = 0.2$, $\zeta^2 = 0.1$, $\epsilon_2 = 10$, $T_1 = 15$, and $T_2 = 10$.}	
\label{fig:detuning}
\end{figure}
		
To further explore the order of magnitude for the thermodynamic quantities during steady state, we illustrate in Fig.~\ref{fig:magnitude}(a) the relationship between $J_{1,ss}$ and system-bath coupling strength $\zeta^2$, with varying subsystem interaction $\lambda$. The figure, plotted on a logarithmic scale, displays a set of parallel lines, each with a slope of unity. From this figure, we can conclude that $J_{1,ss}$ is proportional to $\zeta^2$ while the order of magnitude of $J_{1,ss}$ is significantly lower than that of $\zeta^2$. The dependency between $J_{1,ss}$ and $\lambda^2$ is presented in Fig.~\ref{fig:magnitude}(b). The resulting graph is essentially identical, suggesting that $J_{1,ss}$ is also proportional to $\lambda^2$. These numerical findings align well with the theoretical analysis outlined in Refs.~\cite{Levy_2014, De_Chiara_2018}. Based on the results presented therein, it is straightforward to demonstrate that the relation $J_{1,ss} \propto \lambda^2 \zeta^2$ holds true, under the conditions that both $\lambda\ll 2|\epsilon_1 - \epsilon_2|$ and $ \zeta^2 \ll 2|\epsilon_1 - \epsilon_2|$ are met. Note that the energy gap $\epsilon_i$ and temperature $T_i$ satisfy $\epsilon_1 / \epsilon_2 < T_1 / T_2$ in this scenario, and thus the LME is consistent with the second law according to Fig.~\ref{fig:condition}. However, even under such circumstances, the heat current $J_{1,ss}$ remains a higher-order term of $\zeta^2$. This confirms our previous argument that the LME is in general inappropriate for characterizing the dynamics near the steady state.

\begin{figure}[tb!]
\centering \includegraphics[scale=0.41]{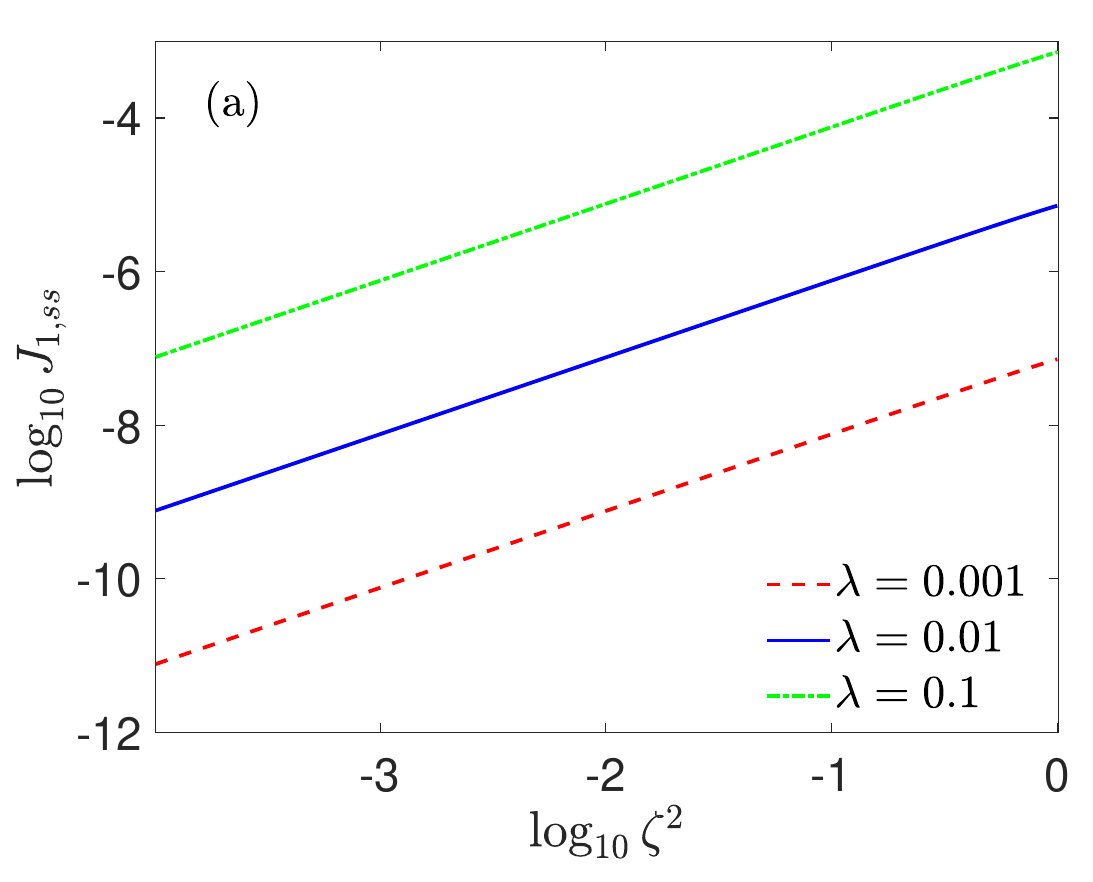}
\includegraphics[scale=0.41]{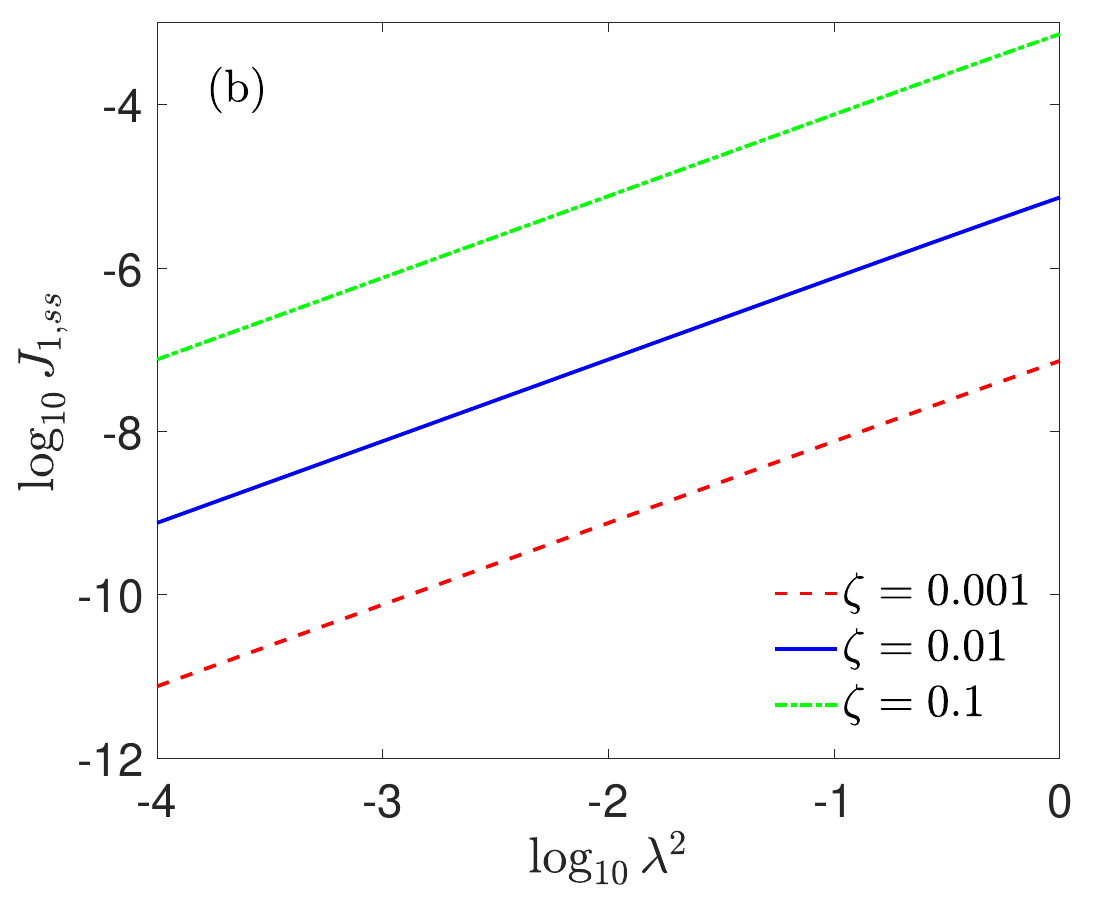}
	
\caption{(a) Steady-state heat current $J_{1,ss}$ as a function of $\zeta^2$, with $\lambda = 0.001$ (red dashed line), $\lambda = 0.01$ (blue solid line), and $\lambda = 0.1$ (green dot dashed line). The order of magnitude of $J_{1,ss}$ is less than that of $\zeta^2$. (b) Steady-state heat current $J_{1,ss}$ as a function of $\lambda^2$, with $\zeta = 0.001$ (red dashed line), $\zeta  = 0.01$ (blue solid line), and $\zeta  = 0.1$ (green dot dashed line). Each of the two subfigures consists of a family of parallel lines, all with a slope of unity. The parameters are set as: $k_B = 1$, $\kappa_1 = \kappa_2 = 10$, $\Omega_1 = \Omega_2 = 1$, $\epsilon_1 = 15$, $\epsilon_2 = 10$, $T_1 = 20$, and $T_2 = 10$.}	
\label{fig:magnitude}
\end{figure}

We now turn to investigate the evolution of the relevant thermodynamic quantities over time. In our numerical simulation,} the energy gaps of the subsystems and the temperatures of the local heat baths in this case are set to be {$\epsilon_1 = 10$, $\epsilon_2 = 5$ and $T_1 = 15$, $T_2 = 10$,} respectively. {Given this choice of the parameters,} the condition $\epsilon_1 /\epsilon_2 > T_1 /T_2$ is satisfied and the second law is violated at the steady state according to the previous result. The strengths of the intrasubsystem interaction $\lambda$ and the system-bath coupling $\zeta$ are taking as {$\lambda = \zeta^2 = 0.5$}, which are much smaller than the energy level difference of the subsystem. The whole system is assumed to be initially prepared at a state whose entropy takes the maximum value, i.e., $\rho_S(0) = \rho_1(0) \otimes  \rho_2(0)$ with $\rho_i(0) = \left( \ket{\uparrow}_i\bra{\uparrow}_i + \ket{\downarrow}_i\bra{\downarrow}_i\right)/2 $. {As depicted in Fig.~\ref{fig:undriven}, the order of $J_i(t)$ at the initial time exceeds that of $\zeta^2$, and the rate of entropy production $\dot{\Sigma}(t)$ remains consistently positive in this region \footnote{Note that both $J_1(t)$ and $J_2(t)$ are negative in the beginning. The reason is that the initial state of the system, the maximum entropy state, is equivalent to an equilibrium state with an infinite effective temperature. If the qubits start from another state, for instance, the ground state, then two heat currents become positive at the initial time.}. However, as the system is close to the steady state (specifically at times $ t \gtrsim 2.23$), the magnitude of $J_i(t)$ diminishes to the order of $10^{-2}$, which becomes a smaller quantity compared to $\zeta^2$. In addition, the entropy production rate $\dot{\Sigma}(t)$ also turns negative, indicating the apparent violation of the second law of thermodynamics. This numerical result reinforces our earlier assertion that the LME works well within its applicable scope, while it may generate the thermodynamic inconsistency near the steady state.} 

{Furthermore, it is also worthwhile to investigate the time $\tau_0$ required for the system to reach a state with $\dot{\Sigma}= 0$, as this time period provides a measure to characterize the time duration in which the LME is consistent with the second law of thermodynamics. Based on our previous discussion, $\tau_0$ is related to the relaxation time $\tau_r$ for the system to reach a steady state. In Appendix \ref{ap:Lyapunov}, we have demonstrated that the thermodynamic observables like $J_i(t)$ can be decomposed into the linear combination of multiple exponentially decaying terms, each characterized by a distinct decay rate. Consequently, it is reasonable to define $\tau_r$ as the time constant corresponding to the term that exhibits the slowest decay rate (see Appendix \ref{ap:Lyapunov} for detail). Figure \ref{fig:crossing} displays how $\tau_0$ and $\tau_r$ vary with respect to $\zeta^2$. The numerical illustration shows that $\tau_r$ is approximately proportional $\zeta^{-2}$ under the present parameter selection, which is consistent with the theoretical analysis (see Appendix \ref{ap:Lyapunov} for detail). In addition, Fig.~\ref{fig:crossing} also demonstrates that, in our numerical example, $\tau_0 \approx 3.8 \tau_r$, which suggests that the time  required for $\dot{\Sigma} < 0$ to occur is typically longer than $3.8 \tau_r$. After evolving for this duration, the system can be considered to have approximately reached the vicinity of a steady state. This result offers the supporting evidence to confirm our argument that the thermodynamic inconsistency can be generated when the system is approaching the steady state.	
	
\begin{figure}[t!]
\centering  \includegraphics[scale=0.415]{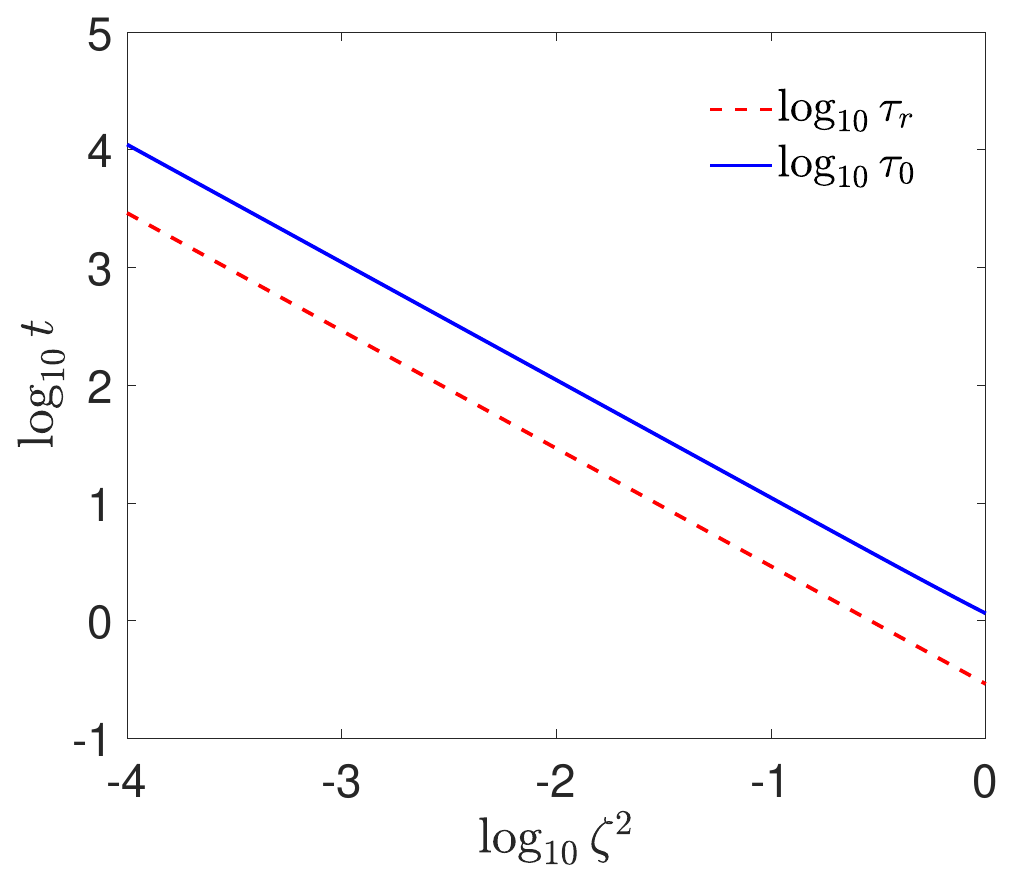}
	
\caption{Moment $\tau_0$ with the zero entropy production rate (blue solid line) and the relaxation time $\tau_r$ (red dashed line) as a function of $\zeta^2$. The system starts from the state of maximum entropy as before. The parameters are set as: $k_B = 1$, $\kappa_1 = \kappa_2 = 10$, $\Omega_1 = \Omega_2 = 1$, $\epsilon_1 = 10$, $\epsilon_2 = 5$, $T_1 = 15$, $T_2 = 10$, and $\lambda = 0.5$.}	
\label{fig:crossing}
\end{figure}

\begin{figure}[t!]
	\centering      
 \includegraphics[scale=0.425]{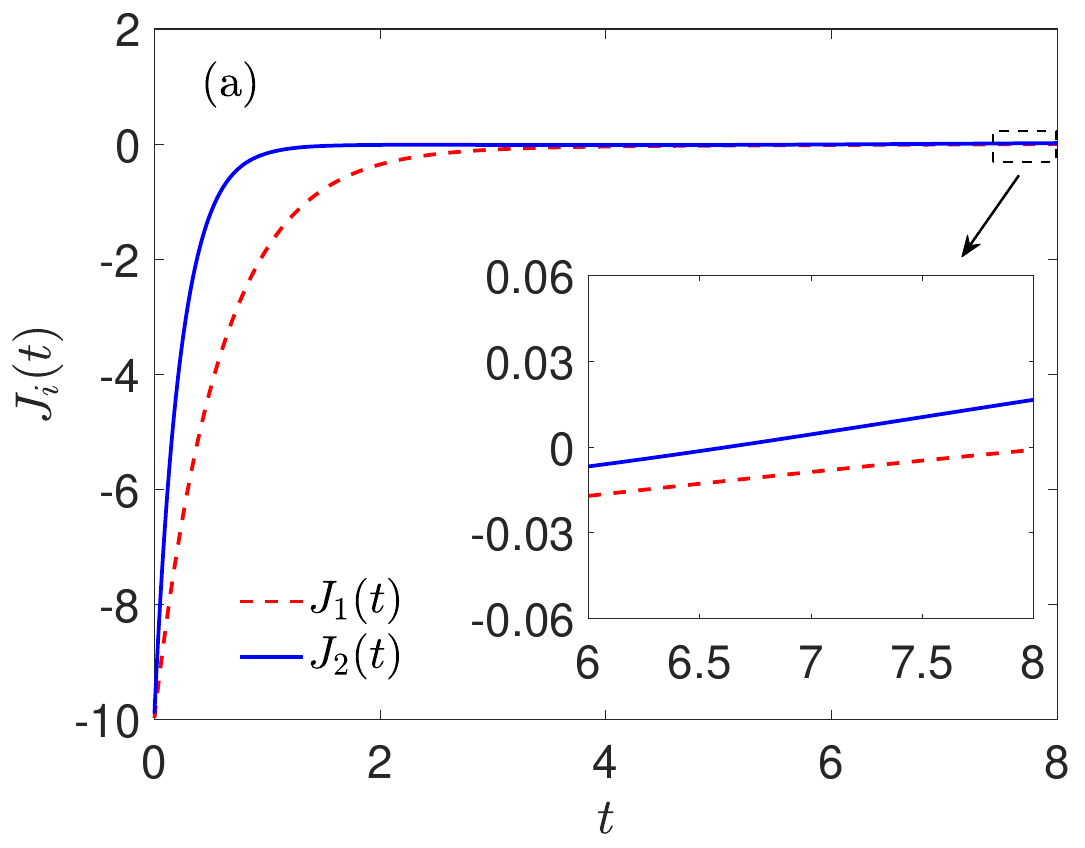}
	
	\centering \includegraphics[scale=0.42]{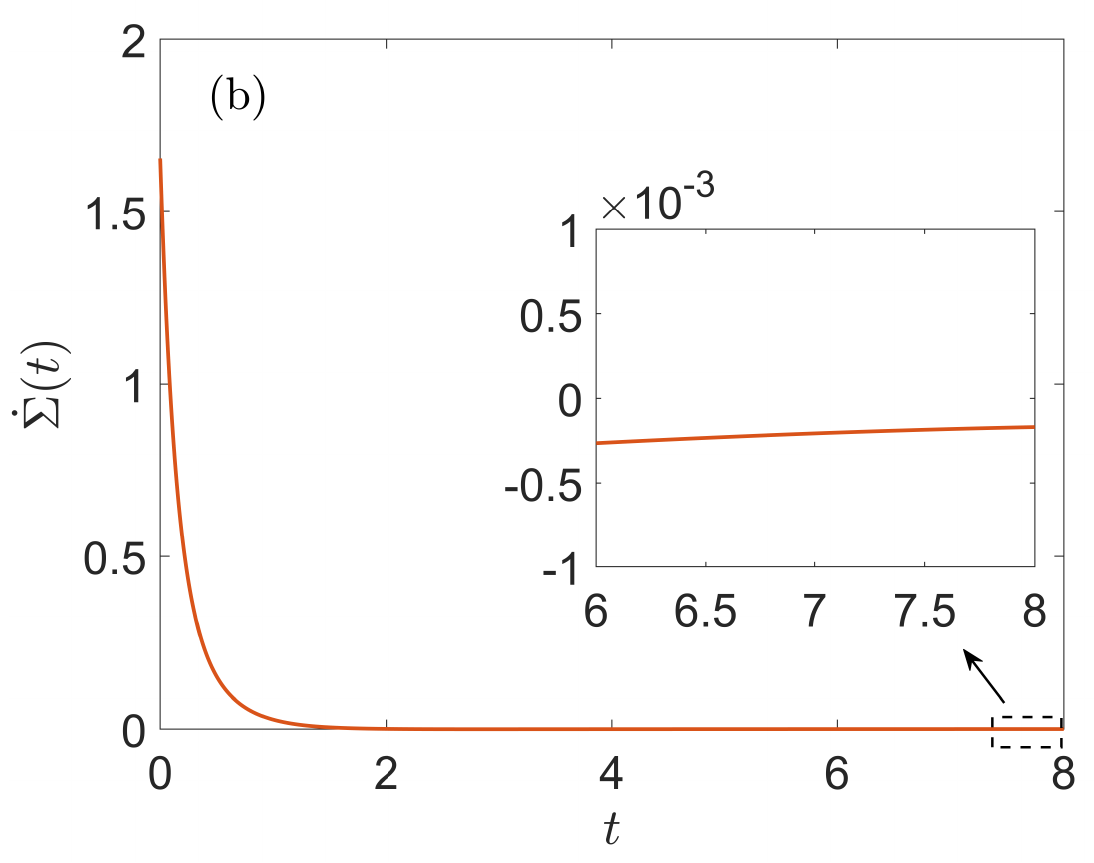}
	\caption{Dynamics of (a) the heat currents $J_1(t)$ (red dashes curve), $J_2(t)$ (blue solid curve), and (b) the entropy production rate $\dot{\Sigma}(t)$ for the system subjected to an external driving. Insets: Similar to the results presented in Fig.~\ref{fig:undriven}, as the system approaches the instantaneous steady state, {it is capable of absorbing heat from the cold bath and emitting heat to the hotter one. In addition, the entropy production rate $\dot{\Sigma} < 0$, indicating that the second law is violated in this region.} Other parameters are set as: $k_B = 1$, $\kappa_1 = \kappa_2 = 10$, $\Omega_1 = \Omega_2 = 1$.}	
	\label{fig:driven}
\end{figure}

Finally, we conclude our numerical demonstration by briefly examining the scenario where the system is subjected to an external driving. However, in contrast to the non-driven case, the system in this situation does not possess a fixed steady state and will usually not  persist in such a state. Furthermore, the temporal evolution of thermodynamic quantities is intimately related to the specific process undergone by the system. Consequently, it is generally challenging to obtain deterministic and universally applicable results in this context. Nevertheless, there are still some predictable conclusions that can be intuitively illustrated through simple examples. By implementing a specific driving protocol, we can validate our earlier finding that the TDLME may also lead to the thermodynamic inconsistencies when the system approaches the instantaneous steady state.} As a simple but typical external driving, here we consider a periodic driving process, $f_i(t) = a_i\sin \omega_i t$, where $a_i$ denotes the strength of the driving Hamiltonian and $\omega_i$ characters the speed of the external driving. {The system starts at the maximum entropy state as well.} Notice that for a two level system, it is always permissible to find the effective temperature $\beta^{\rm eff}_i(t)$ to make sure that Eq.~(\ref{eq:detail}) is satisfied, even if the adiabatic phase $\xi_n$ is nonzero. This advantage allows us to discuss the validity problem of the second thermodynamic law for TDLME. In our numerical analysis, the strengths of the driving are taken as {$a_1 = a_2 = 2$}. {The values of the driving frequencies, $\omega_1$ and $\omega_2$, are determined by the time it takes for the system to approach a near-steady state. As shown in Figs.~\ref{fig:undriven} and \ref{fig:crossing}, we have $\tau_0 \approx 2.23$, and thus the frequencies} are set to be {$\omega_1 = \omega_2 = 0.2$} to guarantee that the external driving is slow enough. In addition, {under this choice of the parameters, we have $\gamma_i^-(t)\approx \gamma_i^{(0)}(2E_{e,i}(t))$ and $\gamma_i^+(t)\approx \gamma_i^{(0)}(-2E_{e,i}(t))$  within the time interval $t \in [0, 8]$, and thus} the following relation holds,
\begin{align}
\frac{\gamma_i^-(t)}{\gamma_i^+(t)} & \approx  \frac{\gamma_i^{(0)}(2E_{e,i}(t))}{\gamma_i^{(0)}(-2E_{e,i}(t))}
\notag
\\
& =  e^{2\beta_i E_{e,i}(t)}.
\end{align}
This expression indicates that the effective temperature can be approximated by $\beta^{\rm eff}_i(t) \approx \beta_i$. The time evolution of the relevant observables for the system under the external driving is displayed in Fig.~\ref{fig:driven}, and the numerical result is similar with that of the non-driven case: {The TDLME does not give rise to the thermodynamic inconsistencies when applied within its appropriate range of application.} The contradiction can only arise when the system is nearing the instantaneous steady state, due to the usage of the inaccurate heat currents.

{\section{conclusion and outlook}
\label{sec:conclusion}
}

In summary, we have generalized the LME to the time-dependent case and derived the TDLME applicable for the boundary-driven system subjected to the external driving. Our derivation is {mainly} based on the the methods developed in Refs.~\cite{RevModPhys.94.045006, PhysRevResearch.3.013165}, and the particular iterative technique as well as the Born-Markov approximation have been employed to simplify the equation. After expanding the phase factors and the jump operators up to the first order of time and performing the secular approximation, we {finally obtain the general form of the TDLME}. Then, we have applied our formalism to the bosonic heat bath case and exactly determined the form of the dissipation rates in the TDLME. Moreover, we have revealed the reason of the thermodynamic inconsistencies generated by the LME and the TDLME via analyzing the order of magnitudes for the {thermodynamic observables}. We have demonstrated that violation of the second thermodynamic law can only appear near the steady state, while this contradiction can be avoided {by applying the equations in their applicable scopes}. Finally, we have presented a toy model to further confirm our results obtained in the paper.
{Our work contributes to clarifying the scope of application for the LME and enhancing the understanding of the evolution of boundary-driven systems.

The dynamics of an open quantum system with a time-dependent Hamiltonian has long been a fascinating and challenging topic. Due to the variety of driving schemes, there is no universal and comprehensive processing framework in this field. The existing methods are often tailored to specific physical scenarios, such as the adiabatic \cite{R_Alicki_1979, PhysRevB.82.134517} or periodic \cite{PhysRevE.55.300, cuetara2015stochastic} driving process. In the present work, we have successfully simplified the Redfield-like equation by assuming that the system Hamiltonian remains relatively unchanged within the correlation time of the bath. However, this assumption inevitably limits the applicability of our formalism in fast driving scenarios, highlighting the urgency of exploring broader solutions. Recently, a theorem based on the global approach is proposed in Ref.~\cite{PhysRevResearch.3.013064}, which provides a new perspective as well as a possible solution for the evolution of fast-driven open quantum systems. In the future, we are looking forward to combining this theorem with the local approach, and to deriving the TDLME applicable to the boundary-driven system under the fast driving condition. Another topic that is not covered in this paper relates to the boundary-driven system with a strong intrasubsystem interaction. According to the analysis presented in Ref.~\cite{Levy_2014}, the LME also gives rise to the thermodynamic inconsistencies for this type of system. However, the reason for this contradiction transcends the scope of the current theoretical framework in this paper. To fully address this issue, further exploration and investigation will be necessary in future research.

}

\section*{acknowledgment}

We thank Prof.~Hui Dong, Dr.~Zhucheng Zhang, {and Dr.~Guitao Lyu} for helpful discussions. This work was supported by the National Natural Science Foundation of China (Grants No.~12088101, No.~U2330401 and No.~12347127) and Science Challenge Project (Grant No.~TZ2018005).

\appendix

{
\section{Comparison of two iterative methods} \label{ap:ite}

Let us first introduce the special iterative technique \cite{RevModPhys.94.045006} employed in the local approach. The differential equation (\ref{eq:Liouville}) can be solved implicitly as,
\begin{align}
\tilde{\rho}(t) = \tilde{\rho}(0) - i\int_{0}^{t}  ds\, 
[\tilde{H}(s), \tilde{\rho}(s)].
\label{eq:integration}
\end{align}
In contrast to the conventional treatment, here we firstly rewrite Eq.~(\ref{eq:Liouville}) as
\begin{align}
\frac{d}{dt} \tilde{\rho}(t) = -i [\tilde{H}_0(t) + \lambda \tilde{H}_I(t), \tilde{\rho}(t)] -i [\zeta \sum_{i=1}^{N}\tilde{V}_i(t), \tilde{\rho}(t)] ,
	\label{eq:14}
\end{align}
and then substitute Eq.~(\ref{eq:integration}) only into the second commutator term of Eq.~(\ref{eq:14}), 
\begin{align}
\frac{d}{dt} \tilde{\rho}(t) = & -i [\tilde{H}_0(t) + \lambda \tilde{H}_I(t), \tilde{\rho}(t)] -i\zeta \sum_{i=1}^{N} [\tilde{V}_i(t), \tilde{\rho}(0)] 
\notag
\\
& - \zeta\sum_{i=1}^{N} \int_{0}^{t}  ds\, [\tilde{V}_i(t),
[\tilde{H}(s), \tilde{\rho}(s)]].
\label{eq:sit}
\end{align}
Taking the partial trace over the baths, we can reach Eq.~(\ref{eq:12}) in the main text. Notice that we do not perform any approximations in the whole iteration process, therefore, Eq.~(\ref{eq:sit}) holds strictly.

Now, we proceed to derive the master equation based on the local approach, but instead utilize the conventional iterative method, which is typically employed in the global approach. Inserting Eq.~(\ref{eq:integration}) into Eq.~(\ref{eq:Liouville}) directly, one arrives at
\begin{align}
\frac{d}{dt} \tilde{\rho}(t) = & -i [\tilde{H}(t), \tilde{\rho}(0)]  -  \int_{0}^{t}  ds\, [\tilde{H}(t),
[\tilde{H}(s), \tilde{\rho}(s)]].
\end{align}
Following the similar procedures in the main text, we firstly take the partial trace over the baths and then make the Born approximation as well as the standard assumption. After all these steps, we obtain
\begin{align}
\frac{d}{dt} \tilde{\rho}_S(t) \approx & -i [\tilde{H}_0(t) + \lambda \tilde{H}_I(t), \tilde{\rho}_S(0)] 
\notag
\\
& - \zeta^2\sum_{i=1}^{N} \int_{0}^{t}  ds \, \Tr_B [\tilde{V}_i(t), [\tilde{V}_i(s), \tilde{\rho}(s)]]
\notag
\\
& -  \int_{0}^{t}  ds \,  [\tilde{H}_0(t) + \lambda \tilde{H}_I(t), [\tilde{H}_0(s) + \lambda \tilde{H}_I(s), \tilde{\rho}_S(s)]].
\label{eq:global}
\end{align}
For time-independent system Hamiltonian, we have $\tilde{H}_0(t) = 0$ and Eq.~(\ref{eq:global}) becomes
\begin{align}
\frac{d}{dt} \tilde{\rho}_S(t) \approx & -i [ \lambda \tilde{H}_I(t), \tilde{\rho}_S(0)] 
\notag
\\
& - \zeta^2\sum_{i=1}^{N} \int_{0}^{t}  ds \, \Tr_B [\tilde{V}_i(t), [\tilde{V}_i(s), \tilde{\rho}(s)]]
\notag
\\
& - \lambda^2 \int_{0}^{t}  ds \,  [ \tilde{H}_I(t), [ \tilde{H}_I(s), \tilde{\rho}_S(s)]].
\end{align}
Given that the equation is only kept up to the second-order infinitesimal, we can approximate it in the following form.
\begin{align}
\frac{d}{dt} \tilde{\rho}_S(t) \approx & -i [ \lambda \tilde{H}_I(t), \tilde{\rho}_S(t)] 
\notag
\\
& - \zeta^2\sum_{i=1}^{N} \int_{0}^{t}  ds \, \Tr_B [\tilde{V}_i(t), [\tilde{V}_i(s), \tilde{\rho}(s)]].
\end{align}
The error of this treatment is $\lambda\zeta^2$, which is a higher-order small quantity \cite{PhysRevE.107.014108}. In this sense, when the system's Hamiltonian is time-independent, the two iterative methods yield the same equation. 

However, when the system is under the external driving, the results are completely different. In this case, $\tilde{H}_0(t) \neq 0$ and thus both the system-bath coupling and the intrasubsystem interaction contribute to the dissipation effect. Furthermore, the Markovian approximation can only be performed on the first integral of Eq.~(\ref{eq:global}). The fundamental reason of the Markovian approximation is that the correlation function of the heat bath decays sufficiently fast over its correlation time, so it is safe to change the lower limit of the integral to be negative infinity. In Eq.~(\ref{eq:global}), however, only the integrand of the first integral is related with the reservoir correlation function [see Eq.~(\ref{eq:markov}) in the main text]. Therefore, it is invalid to make Markovian approximation on the second integral term. In fact, integral result of this term heavily depends on the protocol the system undergoes and it is generally not possible to simplify this integral further without additional assumptions. In other words, the master equation (\ref{eq:global}) derived by the conventional iterative method is typically a complicated integro-differential equation, which is inconvenient for application.

To summarize, when the system is independent of the external driving, the equations derived from these two iterative methods differ by only a third-order small quantity. However, when the system is subjected to the external driving, the results obtained through the special iterative technique are indeed simpler and more convenient in theoretical discussions. In our work, we consider the evolution of the time-dependent boundary-driven system. Thus, the special iterative technique plays an essential role in our derivation.
}

{
\section{The master equation after the Taylor expansion and the rotating wave approximation} 
\label{ap:TDLME}

Inserting Eqs.~(\ref{eq:phi}) and (\ref{eq:a}) into Eq.~(\ref{eq:markov}), we find that $\mathcal{L}_i[\tilde{\rho}_S(t)]$ can be decomposed into two parts,
\begin{align}
\mathcal{L}_i[\tilde{\rho}_S(t)] = \mathcal{L}^{(0)}_i[\tilde{\rho}_S(t)] + \mathcal{L}^{(1)}_i[\tilde{\rho}_S(t)],
\end{align}
with
\begin{widetext}
\begin{align}
\mathcal{L}^{(0)}_i[\tilde{\rho}_S(t)] 
\approx & \sum_{\mu,n,m}  \sum_{\mu',n',m'} e^{i\left[ \phi_{n'm'}(t) - \phi_{nm}(t)\right] } \Gamma^{(0)}_{i\mu'\mu} (\dot{\phi}_{nm}(t)) \left[\mathcal{A}_{i\mu}(n,m,t) \tilde{\rho}_S\mathcal{A}^{\dagger}_{i\mu'}(m',n',t) - \mathcal{A}^{\dagger}_{i\mu'}(m',n',t)\mathcal{A}_{i\mu}(n,m,t) \tilde{\rho}_S\right] + \text{\rm H.c.},
\label{eq:l0}
\\ 
\mathcal{L}^{(1)}_i[\tilde{\rho}_S(t)] 
\approx &  \sum_{\mu,n,m}  \sum_{\mu',n',m'} e^{i\left[ \phi_{n'm'}(t) - \phi_{nm}(t)\right] } \Gamma^{(1)}_{i\mu'\mu} (\dot{\phi}_{nm}(t)) \left[\dot{\mathcal{A}}_{i\mu}(n,m,t) \tilde{\rho}_S\mathcal{A}^{\dagger}_{i\mu'}(m',n',t) - \mathcal{A}^{\dagger}_{i\mu'}(m',n',t)\dot{\mathcal{A}}_{i\mu}(n,m,t) \tilde{\rho}_S\right] 
+ \text{\rm H.c.}.
\label{eq:l1}
\end{align}
\end{widetext}
Here, we have introduced the parameters
\begin{align}
\Gamma^{(0)}_{i\mu'\mu} (\dot{\phi}_{nm}(t)) & \equiv \int_{0}^{+\infty} ds \,  e^{i s\dot{\phi}_{nm}(t)}
C_{i\mu' \mu}(s) 
\\
\Gamma^{(1)}_{i\mu'\mu} (\dot{\phi}_{nm}(t)) & \equiv -\int_{0}^{+\infty} ds \,  e^{i s\dot{\phi}_{nm}(t)}
C_{i\mu' \mu}(s) s.
\end{align}
to denote the zeroth- and first-order dissipation rates, respectively. 

Notice that both $\mathcal{L}_i^{(0)}$ and $\mathcal{L}_i^{(1)}$ contain the summation over six indices, which is extremely cumbersome for the application. In order to reduce the complexity, we apply the secular rotating wave approximation to eliminate the fast oscillating terms. According to the statement in Sec.~\ref{sec:derivation}, only the terms with $n = m$, $n' = m'$ and the terms with $n = n'$ and $m = m'$ are kept. Therefore, $\mathcal{L}_i^{(0)}$ and $\mathcal{L}^{(1)}_i$ can be simplified to the following form of the quadruple summations,
\begin{widetext}
\begin{align}
\mathcal{L}^{(0)}_i[\tilde{\rho}_S(t)]  \equiv &  
\sum_{\mu,\mu'}  \sum_{n,n'} \Gamma^{(0)}_{i\mu'\mu} (0) \left[{\mathcal{A}}_{i\mu}(n,n,t) \tilde{\rho}_S\mathcal{A}^{\dagger}_{i\mu'}(n',n',t) - \mathcal{A}^{\dagger}_{i\mu'}(n',n',t) {\mathcal{A}}_{i\mu}(n,n,t) \tilde{\rho}_S\right] 
\notag
\\ 
& +  \sum_{\mu,\mu'}  \sum_{n\neq m} \Gamma^{(0)}_{i\mu'\mu} (\dot{\phi}_{nm}(t)) \left[{\mathcal{A}}_{i\mu}(n,m,t) \tilde{\rho}_S\mathcal{A}^{\dagger}_{i\mu'}(m,n,t) - \mathcal{A}^{\dagger}_{i\mu'}(m,n,t){\mathcal{A}}_{i\mu}(n,m,t) \tilde{\rho}_S\right] 
+ \text{\rm H.c.},
\label{eq:li0}
\\
\mathcal{L}^{(1)}_i[\tilde{\rho}_S(t)]  \equiv &  
\sum_{\mu, \mu'}  \sum_{n,n'} \Gamma^{(1)}_{i\mu'\mu} (0) \left[\dot{\mathcal{A}}_{i\mu}(n,n,t) \tilde{\rho}_S\mathcal{A}^{\dagger}_{i\mu'}(n',n',t) - \mathcal{A}^{\dagger}_{i\mu'}(n',n',t)\dot{\mathcal{A}}_{i\mu}(n,n,t) \tilde{\rho}_S\right] 
\notag
\\ 
& +  \sum_{\mu, \mu'}  \sum_{n\neq m} \Gamma^{(1)}_{i\mu'\mu} (\dot{\phi}_{nm}(t)) \left[\dot{\mathcal{A}}_{i\mu}(n,m,t) \tilde{\rho}_S\mathcal{A}^{\dagger}_{i\mu'}(m,n,t) - \mathcal{A}^{\dagger}_{i\mu'}(m,n,t)\dot{\mathcal{A}}_{i\mu}(n,m,t) \tilde{\rho}_S\right] 
+ \text{\rm H.c.}.
\label{eq:li1}
\end{align}
\end{widetext} 
In the literature, it is more typical to reformulate $\mathcal{L}^{(0)}_i$ into the LGKS form \cite{HMW, KJ, HPBFP, lindblad1976generators, gorini1976completely, carmichael2013statistical}. By introducing the parameters
\begin{align}
&\gamma^{(0)}_{i\mu'\mu} (\dot{\phi}_{nm}) \equiv \Gamma^{(0)}_{i\mu'\mu} + \left( \Gamma^{(0)}_{i\mu\mu'}\right) ^*,
\\
& S^{(0)}_{i\mu'\mu} (\dot{\phi}_{nm})\equiv \frac{1}{2i}\left[ \Gamma^{(0)}_{i\mu'\mu} - \left( \Gamma^{(0)}_{i\mu \mu'}\right) ^*\right], 
\end{align}
$\mathcal{L}^{(0)}_i[\tilde{\rho}_S(t)]$ can be equivalently written as,
\begin{align}
\mathcal{L}^{(0)}_i[\tilde{\rho}_S(t)] = & -i[H^{(0)}_{i,LS}, \tilde{\rho}_S(t)] + \mathcal{D}_i^{(0)}[\tilde{\rho}_S(t)],
\end{align}
with Lamb shift Hamiltonian $H^{(0)}_{i,LS}$ and LGKS form dissipator $\mathcal{D}_i^{(0)}[\tilde{\rho}_S(t)]$ being}
\begin{widetext}
\begin{align}
H^{(0)}_{i,LS} \equiv & \sum_{\mu,\mu'}  \sum_{n,n'} S^{(0)}_{i\mu'\mu}(0) \mathcal{A}^{\dagger}_{i\mu'}(n',n',t) \mathcal{A}_{i\mu}(n,n,t)
+ \sum_{\mu,\mu'}  \sum_{n\neq m} S^{(0)}_{i\mu'\mu}(\dot{\phi}_{nm}) \mathcal{A}^{\dagger}_{i\mu'}(m,n,t) \mathcal{A}_{i\mu}(n,m,t)
\label{eq:lamb}
\end{align}
and
\begin{align}
\mathcal{D}_i^{(0)}[\tilde{\rho}_S(t)] \equiv &
\sum_{\mu,\mu'}  \sum_{n, n'}  \gamma^{(0)}_{i\mu'\mu}(0)
\left[ \mathcal{A}_{i\mu}(n,n,t) \tilde{\rho}_S\mathcal{A}^{\dagger}_{i\mu'}(n',n',t) - \frac{1}{2}\left\lbrace \mathcal{A}^{\dagger}_{i\mu'}(n',n',t)\mathcal{A}_{i\mu}(n,n,t),
\tilde{\rho}_S\right\rbrace \right]
\notag
\\
& + \sum_{\mu,\mu'}  \sum_{n \neq m}  \gamma^{(0)}_{i\mu'\mu}(\dot{\phi}_{nm})
\left[ \mathcal{A}_{i\mu}(n,m,t) \tilde{\rho}_S\mathcal{A}^{\dagger}_{i\mu'}(m,n,t) 
- \frac{1}{2}\left\lbrace \mathcal{A}^{\dagger}_{i\mu'}(m,n,t)\mathcal{A}_{i\mu}(n,m,t),
\tilde{\rho}_S\right\rbrace \right].
\label{eq:d0}
\end{align}
\end{widetext}

\section{Recovery of the LME} \label{ap:recovery}

Considering a time-independent bare Hamiltonian $H_0$. In this case, both the eigenvalue $E_n(t)$ and the eigenvector $\ket{n(t)}$ are time-independent, thus there is no ambiguity to abbreviate them as $E_n$ and $\ket{n}$, respectively. The transformation operator $U_S(t)$ becomes
\begin{align}
U_S(t) = \sum_{n} e^{-i E_n t}\ket{n} \bra{n} =  e^{-i H_0 t}.
\end{align}
In addition, the first order correction term $\mathcal{L}_i^{(1)}[\tilde{\rho}_S(t)]$ in Eq.~(\ref{eq:TDLME}) vanishes, and the TDLME can be reduced into 
\begin{align}
\frac{d}{dt} \tilde{\rho}_S = &  -i \left[ \tilde{H}_0(t) + \lambda\tilde{H}_I(t), \tilde{\rho}_S(t)\right]  
+ \zeta^2 \sum_{i = 1}^{N}  \mathcal{D}_i^{(0)}[\tilde{\rho}_S(t)]. 
\end{align}
According to the inverse transformation relation $\rho_S(t) = U_S(t) \tilde{\rho}_S(t) U^{\dagger}_S(t) $, the TDLME of $\rho_S(t)$ reads,
\begin{align}
\frac{d}{dt}\rho_S & = 
\frac{d U_S}{dt} \tilde{\rho}_S U^{\dagger}_S +  U_S \frac{d \tilde{\rho}_S }{dt}  U_S^{\dagger} + U_S  \tilde{\rho}_S \frac{d U_S^{\dagger} }{dt}
\notag
\\
& = \frac{d U_S}{dt} U_S^{\dagger} \rho_S + \rho_S  U_S\frac{d U_S^{\dagger}}{dt} +  U_S \frac{d \tilde{\rho}_S }{dt}  U_S^{\dagger}
\notag
\\
& =  - i \left[ i \frac{d U_S}{dt} U_S^{\dagger} , \rho_S  \right]  +  U_S \frac{d \tilde{\rho}_S }{dt}  U_S^{\dagger}. 
\end{align}
In the last line, we have used the identity $d \left(  U_S U_S^{\dagger}  \right) = \left(d U_S\right) U_S^{\dagger} +  U_S d U^{\dagger}_S  = 0 $. Note that
\begin{align}
& - i \left[ i \frac{d U_S}{dt} U_S^{\dagger} , \rho_S  \right]  +  U_S  \left(  -i \left[ \tilde{H}_0 + \lambda\tilde{H}_I, \tilde{\rho}_S\right]  \right)     U_S^{\dagger}
\notag
\\
= & - i \left[ H_0 + \lambda H_I, \rho_S  \right],
\end{align}
thus we only need to calculate $ U_S \mathcal{D}_i^{(0)}[\tilde{\rho}_S(t)] U_S^{\dagger}$. Without loss of generality, here we just present the evaluation of the term $U_S  \mathcal{A}_{i\mu}(n,m) \tilde{\rho}_S(t)\mathcal{A}^{\dagger}_{i\mu'}(m,n) U_S^{\dagger}$, and the calculation of the other terms in $ U_S \mathcal{D}_i^{(0)}[\tilde{\rho}_S(t)] U_S^{\dagger}$ can be performed by following
the same procedure. From Eq.~(\ref{eq:aadagger}), we can see
\begin{align}
U_S \mathcal{A}_{i\mu}(n,m) U_S^{\dagger}  = &\bra{n} A_{i\mu} \ket{m} U_S\ket{n} \bra{m}U_S^{\dagger} 
\notag
\\
= & e^{-i\left(E_n - E_m \right)t } \mathcal{A}_{i\mu}(n,m)
\end{align}
and
\begin{align}
U_S \mathcal{A}^{\dagger}_{i\mu}(m,n) U_S^{\dagger}  = & \left[  U_S \mathcal{A}_{i\mu}(n,m) U_S^{\dagger} \right] ^{\dagger} 
\notag
\\
= & e^{-i\left(E_n - E_m \right)t } \mathcal{A}_{i\mu}^{\dagger}(m,n).
\end{align}
The substitution of them into $U_S\mathcal{A}_{i\mu}(n,m) \tilde{\rho}_S \mathcal{A}^{\dagger}_{i\mu'}(m,n) U_S^{\dagger}$ leads to
\begin{align}
& U_S\mathcal{A}_{i\mu}(n,m) \tilde{\rho}_S(t)\mathcal{A}^{\dagger}_{i\mu'}(m,n) U_S^{\dagger}
\notag
\\
= & \mathcal{A}_{i\mu}(n,m) {\rho}_S(t)\mathcal{A}^{\dagger}_{i\mu'}(m,n).
\end{align}
Following the similar computation, we can prove
\begin{align}
U_S \mathcal{D}_i^{(0)}[\tilde{\rho}_S(t)] U_S^{\dagger} = \mathcal{D}_i^{(0)}[{\rho}_S(t)]. 
\end{align}
Therefore, the TDLME (\ref{eq:TDLME}) in the Schr\"{o}dinger picture recovers to
\begin{align}
\frac{d}{dt}\rho_S =  -i \left[  H_0 + \lambda  {H}_I, \rho_S(t) \right]  
+ \zeta^2 \sum_{i = 1}^{N}  \mathcal{D}_i^{(0)}[\rho_S(t) ],
\end{align}
which takes the form of the LME.

\section{The calculation of bath correlation function} 
\label{ap:correlation}

The combination of Eqs.~(\ref{eq:cf}) and (\ref{eq:tildeBi}) in the main text yields
\begin{align}
&C_i(s)
\notag
\\
= & \Tr\left[\tilde{B}^{\dagger}_i(s) \tilde{B}_i(0) \rho_{B_i}\right] 
\notag
\\
= & \sum_{k_{i},k_{i}'} g_{k_{i}} g_{k_{i}'} \Tr\left[ \left( b_{k_{i}}^{\dagger} e^{i\omega_{k_{i}} s} + b_{k_{i}} e^{-i\omega_{k_{i}} s} \right)
\left( b_{k_{i}'}^{\dagger} + b_{k_{i}'}  \right) \rho_{B_i} \right] 
\notag
\\
= & \sum_{k_{i}} g^2_{k_{i}}  \Tr\left[ \left( b_{k_{i}}^{\dagger} b_{k_{i}} e^{i\omega_{k_{i}} s} + b_{k_{i}} b_{k_{i}}^{\dagger} e^{-i\omega_{k_{i}} s} \right) \rho_{B_i} \right].
\end{align}
Note that
\begin{align}
\Tr\left[ b_{k_{i}}^{\dagger} b_{k_{i}}  \rho_{B_i} \right]  = \frac{1}{e^{\beta_i \omega_{k_{i}}} -1},
\\
\Tr\left[ b_{k_{i}} b_{k_{i}}^{\dagger}  \rho_{B_i} \right]
= \frac{e^{\beta_i \omega_{k_{i}}}}{e^{\beta_i \omega_{k_{i}}} -1},
\end{align}
so the correlation function takes the form,
\begin{align}
C_i(s) & = \sum_{k} g^2_{k_{i}} \left( \frac{e^{i \omega_{k_{i}}s} }{e^{\beta_i \omega_{k_{i}}} -1}  + \frac{e^{\beta_i \omega_{k_{i}}} e^{-i \omega_{k_{i}}s}}{e^{\beta_i \omega_{k_{i}}} -1}\right) 
\notag
\\
& =  \sum_{k_{i}} g^2_{k_{i}} \left[  \frac{e^{\beta_i \omega_{k_{i}}} + 1 }{e^{\beta_i \omega_{k_{i}}} -1} \cos(\omega_{k_{i}} s) - i \sin(\omega_{k_{i}} s) \right]
\notag
\\
& =  \sum_{k_{i}} g^2_{k_{i}} \left[  \coth\left( \frac{\beta_i\omega_{k_{i}}}{2}\right)  \cos(\omega_{k_{i}} s) - i \sin(\omega_{k_{i}} s) \right] . 
\end{align}
After taking the continuum limit for the field frequency $\omega_{k_{i}}$ and coupling strength $g_{k_{i}}$,
\begin{align}
& \sum_{k_{i}}  \to \int_{0}^{+\infty} d\omega \rho_i(\omega),
\\
& g_{k_{i}} \to g_i(\omega),
\end{align}
we finally reach the result presented in the main text,
\begin{align}
&C_i(s) 
\notag
\\
= & \int_{0}^{+\infty}d \omega \,  \rho_i(\omega)g_i^2(\omega)
\left[ \coth\left(\frac{\beta_i \omega}{2} \right)\cos\left(\omega s \right) -i\sin\left(\omega s \right) \right].
\end{align}

\section{The calculation of parameters $\gamma^{(0)}_{i}$ and $S^{(0)}_{i}$} 
\label{ap:parameter}

Let us firstly discuss $\gamma^{(0)}_{i}(\dot{\phi})$. According to Eq.~(\ref{eq:gamma0i}) in the main text, we have
\begin{widetext}
\begin{align}
\gamma^{(0)}_{i}(\dot{\phi}) & =  \int_{0}^{+\infty}d \omega \int_{0}^{+\infty}ds \, \mathcal{J}_i(\omega) \left\lbrace  \coth\left(\frac{\beta_i \omega}{2} \right)\left[ \cos\left(\omega +\dot{\phi} \right)s + \cos\left(\omega- \dot{\phi} \right)s \right]  +\left[ \cos\left(\omega -\dot{\phi} \right)s - \cos\left(\omega + \dot{\phi} \right)s \right] \right\rbrace 
\notag
\\
& = \pi\int_{0}^{+\infty}d \omega \, \mathcal{J}_i(\omega)  \left[ \coth\left(\frac{\beta_i \omega}{2}  \right) + 1\right] \delta\left( \omega - \dot{\phi}\right) + \mathcal{J}_i(\omega)  \left[ \coth\left(\frac{\beta_i \omega}{2}  \right) - 1\right] \delta\left( \omega + \dot{\phi}\right)
\notag
\\
& = \pi\left\lbrace  \mathcal{J}_i(\dot{\phi})  \left[ \coth\left(\frac{\beta_i \dot{\phi}}{2}  \right) + 1\right] u\left( \dot{\phi}\right) - \mathcal{J}_i(-\dot{\phi})  \left[ \coth\left(\frac{\beta_i \dot{\phi}}{2}  \right) + 1\right] u\left( -\dot{\phi}\right)\right\rbrace 
\notag
\\
& = \pi \mathcal{J}_i(\dot{\phi})  \left[ \coth\left(\frac{\beta_i \dot{\phi}}{2}  \right) + 1\right] \left[ u\left(\dot{\phi} \right) + u\left( -\dot{\phi}\right) \right] 
\notag
\\
& = \pi  \mathcal{J}_i(\dot{\phi})  \left[ \coth\left(\frac{\beta_i \dot{\phi}}{2}  \right) + 1\right].
\end{align}
\end{widetext}
Here,
\begin{align}
u(x) = \begin{cases}
0, (x< 0)
\\
1,(x> 0)
\end{cases}
\end{align}
is a step function. Note that $\gamma^{(0)}_{i}(0)$ is defined as
\begin{align}
\gamma^{(0)}_{i}(0) & = \lim_{x \to 0} \gamma^{(0)}_{i}(x)
\notag
\\
& =  4 \kappa_i k_B T_i.
\end{align}

We then turn to calculate $S^{(0)}_{i}$. For an Ohmic spectral density defined by Eq.~(\ref{eq:spectral}), the real and imaginary parts of correlation function $C_i(s)$ can be obtained analytically as \cite{HPBFP}
\begin{align}
& \int_{0}^{+\infty}d \omega \,  \mathcal{J}_i(\omega)
\coth\left(\frac{\beta_i \omega}{2} \right)\cos\left(\omega s \right) 
\notag
\\
= & \frac{2\kappa_i \Omega_i^2}{\beta_i} \sum_{n= - \infty}^{+ \infty} \frac{\Omega_i e^{-\Omega_i|s|} - |\nu_{n,i}|  e^{-|\nu_{n,i} s|} }{\Omega_i^2 - \nu_{n,i}^2},
\label{eq:realpart}
\\
 & \int_{0}^{+\infty}d \omega \,  \mathcal{J}_i(\omega)
\sin\left(\omega s \right)
= \kappa_i \Omega_i^2 e^{-\Omega_i|s|} {\rm sgn} (s).
\label{eq:imaginary}
\end{align}
Here, $\nu_{n,i} \equiv 2\pi n k_B T_i$ is Matsubara frequency and ${\rm sgn}(x)$ is known as sign function. In the high temperature limit $k_B T_i \gtrsim \Omega_i$, the terms for which $n \neq 0$ in Eq.~(\ref{eq:realpart}) decay to zero, therefore, the real part of $C_i(s)$ can be approximated by \cite{HPBFP}
\begin{align}
\int_{0}^{+\infty}d \omega \,  \mathcal{J}_i(\omega)
\coth\left(\frac{\beta_i \omega}{2} \right)\cos\left(\omega s \right) 
\approx \frac{2\kappa_i \Omega_i}{\beta_i} e^{-\Omega_i|s|}.
\label{eq:real}
\end{align}
Inserting Eqs.~(\ref{eq:imaginary}) and (\ref{eq:real}) into Eq.~(\ref{eq:s0i}) in the main text, we find

\begin{align}
S^{(0)}_{i} \approx & \int_{0}^{+\infty} ds \frac{2\kappa_i \Omega_i}{\beta_i} e^{-\Omega_is} \sin(\dot{\phi}_i s) - \kappa_i \Omega_i^2 e^{-\Omega_is} \cos(\dot{\phi}_i s)
\notag
\\
= & \frac{2\kappa_i \Omega_i}{\beta_i} \frac{\dot{\phi}_i}{\dot{\phi}_i^2 + \Omega_i^2} - \kappa_i \Omega_i^2 \frac{\Omega_i}{\dot{\phi}_i^2 + \Omega_i^2}  
\notag
\\
= & \frac{\kappa_i \Omega_i}{\dot{\phi}_i^2 + \Omega_i^2} \left( 2k_B T_i \dot{\phi}_i - \Omega_i^2 \right).
\end{align}

\section{A part of subsystems contacting with \\ the local heat baths} 
\label{ap:parts}

\begin{figure}[t!]
	
\centering \includegraphics[scale=0.43]{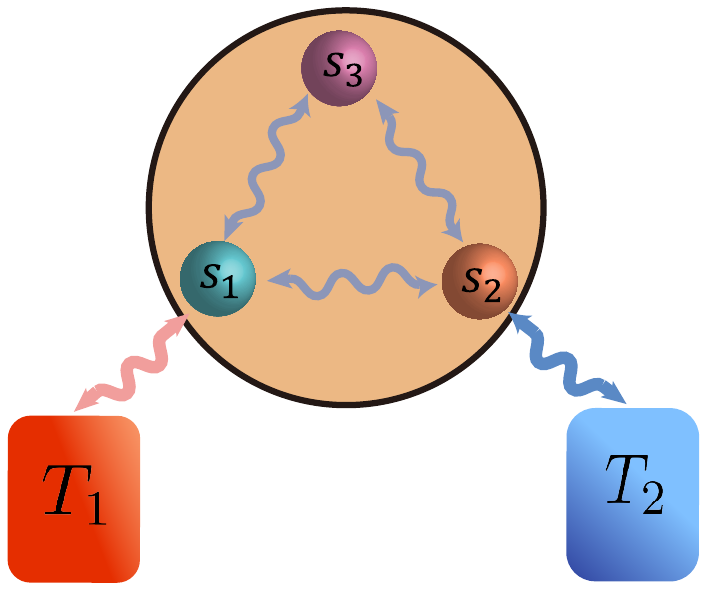}
	
\caption{Schematic picture for the demonstration system. The whole system is made up of three subsystems, $s_1$, $s_2$, and $s_3$. The subsystems interact with each other, but only two of them, $s_1$ and $s_2$, are connected with the local heat baths.}	
\label{fig:part}
\end{figure}

For simplicity, here we consider the situation where the system of interest consists of three subsystems, and it is intuitive to extend the conclusion to the general case. As shown in Fig.~\ref{fig:part}, the system is composed of three parts, $s_1$, $s_2$, and $s_3$. The subsystems interact weakly with each other but only $s_1$ and $s_2$ are connected to their own local heat baths. The evolution of the whole system are described by the LME,
\begin{align}
	\frac{d}{dt}\rho_S = &  -i \left[ \sum_{i = 1}^3 H_{s_i} + \lambda  {H}_I , \rho_S(t) \right]  + \zeta^2 \sum_{i = 1}^2 \mathcal{D}_i [\rho_S(t) ].
	\label{eq:lme123}
\end{align}
The definition of the notations in this equation has already been given in Sec.~\ref{sec:condition}, so we will not reiterate them here.

The key point of the proof is to find the proper approximate expression for the density operator near the steady state. To achieve this goal, we need to firstly derive the master equation for each subsystem. Let us focus on $s_1$ as an example, and the master equation for the other subsystems can be obtained by following the same procedure. According to the theory of quantum mechanics, any operator can be decomposed into a linear combination of tensor products of the eigenoperators pertaining to the different Hilbert
subspaces. Therefore, we can rewrite the density operator $\rho_S$ into
\begin{align}
	\mathcal{\rho}_S & = \sum_{\alpha_1,\alpha_2,\alpha_3} 
	c_{\alpha_1\alpha_2\alpha_3} L_{\alpha_1} \otimes L_{\alpha_2} \otimes L_{\alpha_3}.
\end{align}
Here, we have adopted the notation $\alpha_i = \{n_i, m_i\}$ for short. After introducing the notation $\Tr_{ij} = \Tr_{i} \Tr_{j} = \Tr_{j} \Tr_{i} $ to denote the partial trace over subsystems $s_i$ and $s_j$, the reduced density operator for $s_1$ can be written as
\begin{align}
	\rho_{s_1}  = \sum_{\alpha_1} \Tr_{23}\left( \sum_{\alpha_2,\alpha_3} c_{\alpha_1\alpha_2\alpha_3} L_{\alpha_2} L_{\alpha_3} \right) 
	L_{\alpha_1}.
\end{align}
Notice that 
\begin{widetext}
	\begin{align}
		\Tr_{23} \left( \left[ H_{s_1}, \rho_S(t) \right]\right)
		= & \sum_{\alpha_1,\alpha_2,\alpha_3} 
		c_{\alpha_1\alpha_2\alpha_3}  \left[H_{s_1}, L_{\alpha_1} \right] \otimes \Tr_{23} \left( L_{\alpha_2} L_{\alpha_3} \right) 
		\notag
		\\
		= & \left[  H_{s_1}, \sum_{\alpha_1} \Tr_{23}\left( \sum_{\alpha_2,\alpha_3} c_{\alpha_1\alpha_2\alpha_3} L_{\alpha_2} L_{\alpha_3} \right) 
		L_{\alpha_1} \right] 
		\notag
		\\
		= & \left[  H_{s_1}, \rho_{s_1} \right] ,
		\\
		\Tr_{23} \left( \left[ H_{s_2} + H_{s_3}, \rho_S(t) \right]\right)
		= & \sum_{\alpha_1,\alpha_2,\alpha_3} 
		c_{\alpha_1\alpha_2\alpha_3}  L_{\alpha_1}\otimes \Tr_{2} \left(\left[ H_{s_2} , L_{\alpha_2} \right] \right) \otimes \Tr_{3} L_{\alpha_3} 
		+  \sum_{\alpha_1,\alpha_2,\alpha_3} 
		c_{\alpha_1\alpha_2\alpha_3}  L_{\alpha_1}\otimes \Tr_{2} L_{\alpha_2} \otimes \Tr_{3} \left(\left[ H_{s_3} , L_{\alpha_3} \right] \right) 
		\notag
		\\
		= & 0,
		\\
		\Tr_{23} \left( \mathcal{D}_1 [\rho_S(t) ] \right) 
		= &  \sum_{\alpha_1} \sum_{\alpha'_1,\alpha'_2,\alpha'_3} \gamma_{\alpha_1} c_{\alpha'_1\alpha'_2\alpha'_3} \left( L_{\alpha_1}  L_{\alpha'_1} L_{\alpha_1}^{\dagger} - \frac{1}{2}\left\lbrace L_{\alpha_1}^{\dagger}L_{\alpha_1}, L_{\alpha'_1} \right\rbrace \right)  \otimes \Tr_{23} \left(  L_{\alpha'_2} L_{\alpha'_3} \right) 
		\notag
		\\
		= & \sum_{\alpha_1} \gamma_{\alpha_1}  L_{\alpha_1}  \left[ \sum_{\alpha'_1}\Tr_{23} \left( \sum_{\alpha'_2,\alpha'_3} c_{\alpha'_1\alpha'_2\alpha'_3}  L_{\alpha'_2} L_{\alpha'_3} \right) L_{\alpha'_1} \right]  L_{\alpha_1}^{\dagger} 
		\notag
		\\
		& - \frac{1}{2} \sum_{\alpha_1} \gamma_{\alpha_1} \left\lbrace L_{\alpha_1}^{\dagger}L_{\alpha_1},  \sum_{\alpha'_1}\Tr_{23} \left( \sum_{\alpha'_2,\alpha'_3} c_{\alpha'_1\alpha'_2\alpha'_3}  L_{\alpha'_2} L_{\alpha'_3} \right) L_{\alpha'_1} \right\rbrace 
		\notag
		\\
		= & \sum_{\alpha_1} \gamma_{\alpha_1} \left( L_{\alpha_1} \rho_{s_1} L_{\alpha_1}^{\dagger} - \frac{1}{2}\left\lbrace L_{\alpha_1}^{\dagger}L_{\alpha_1}, \rho_{s_1} \right\rbrace \right)  
		\notag
		\\
		= & \mathcal{D}_1 [\rho_{s_1} ],
	\end{align}
	\begin{align}
		\Tr_{23} \left( \mathcal{D}_2 [\rho_S(t) ] \right) = &  \sum_{\alpha_2} \sum_{\alpha'_1,\alpha'_2,\alpha'_3} \gamma_{\alpha_2} c_{\alpha'_1\alpha'_2\alpha'_3} L_{\alpha'_1} \otimes \Tr_{2} \left( L_{\alpha_2}  L_{\alpha'_2} L_{\alpha_2}^{\dagger} - \frac{1}{2}\left\lbrace L_{\alpha_2}^{\dagger}L_{\alpha_2}, L_{\alpha'_2} \right\rbrace \right) \otimes \Tr_3 L_{\alpha'_3}
		\notag
		\\
		= & 0,
	\end{align}
\end{widetext}
therefore, the master equation for $\rho_{s_1}$ reads,
\begin{align}
	\frac{d}{dt}\rho_{s_1} = &  -i \left[ H_{s_1} , \rho_{s_1} \right] -i \lambda \Tr_{23} \left(\left[ H_I, \rho_S \right]  \right)   + \zeta^2 \mathcal{D}_1 [\rho_{s_1}(t) ].
\end{align}
Similarly, we can prove that the LMEs for $s_2$ and $s_3$ take the form,
\begin{align}
	\frac{d}{dt}\rho_{s_2} = &  -i \left[ H_{s_2} , \rho_{s_2} \right] -i \lambda \Tr_{13} \left(\left[ H_I, \rho_S \right]  \right)   + \zeta^2 \mathcal{D}_2 [\rho_{s_2}(t) ],
	\\
	\frac{d}{dt}\rho_{s_3} = &  -i \left[ H_{s_3} , \rho_{s_3} \right] -i \lambda \Tr_{12} \left(\left[ H_I, \rho_S \right]  \right).
\end{align}
The steady state $\rho_{s_i,ss}(\lambda)$ for subsystem $s_i$ is exactly determined by these three master equations. According to the analysis in Sec.~\ref{sec:condition}, if we denote the steady state for the whole system by $\rho_{ss}(\lambda)$ and further suppose that $\rho_{ss}(\lambda)$ is a smooth function of the parameter $\lambda$, then $\rho_{ss}(\lambda)$ can be approximated by its first order Taylor expansion as follows, 
\begin{align}
	\rho_{ss}(\lambda) & \approx \rho_{ss}(0) + \lambda \rho^{(1)}
	\notag
	\\
	& =  \otimes \prod_{i = 1}^3 \rho_{s_i,ss}(0)+ \lambda \rho^{(1)}.
\end{align}
Here, $\rho^{(1)}$ represents the first order correction term. Therefore, the heat current of $s_1$ at the steady state becomes
\begin{align}
J_{s_1,ss} \equiv & \zeta^2\Tr\left\lbrace \left(\sum_{i = 1}^3 H_{s_i} + \lambda H_I\right) \mathcal{D}_1 [\rho_{ss}(\lambda) ] \right\rbrace 
\notag
\\
= & \zeta^2\Tr\left\lbrace \left(\sum_{i = 1}^3 H_{s_i} + \lambda H_I\right) \left( \mathcal{D}_1 [\rho_{ss}(0)] + \lambda  \mathcal{D}_1 [\rho^{(1)}] \right)  \right\rbrace 
\notag
\\
= & \lambda\zeta^2 \Tr \left( \sum_{i = 1}^3 H_{s_i}\mathcal{D}_1 [\rho^{(1)}] + H_I \mathcal{D}_1 [\rho_{ss}(0)]\right) + O(\lambda^2\zeta^2).
\end{align}
The heat flow of $s_2$ can also be calculated by using a similar method, and the result is analogous to that of $J_{s_1,ss} $. In summary, the leading term of the heat currents is still of the order of $\lambda\zeta^2 $.

{
\section{Exploring the thermodynamic properties of the Gaussian system} 
\label{ap:Lyapunov}

For a Gaussian system, its dynamics is completely characterized by $2\times2$ covariance matrix $C$, whose element $C_{ij}$ is defined as \cite{RevModPhys.94.045006, Levy_2014, vcapek2002zeroth}
\begin{align}
C_{ij} & \equiv \Tr\left( \sigma^{\dagger}_i \sigma_j \rho_S\right) 
\notag
\\
& = \left\langle \sigma^{\dagger}_i \sigma_j  \right\rangle .
\end{align}
By introducing the drift matrix
\begin{widetext}
\begin{align}
W = 
\begin{pmatrix}
-\frac{\zeta^2}{2}\left( \gamma_1^+ + \gamma_1^-\right) + i\left( \epsilon_1 - \epsilon_2\right)   & i \lambda
\\
i\lambda & -\frac{\zeta^2}{2}\left( \gamma_2^+ + \gamma_2^-\right) - i\left( \epsilon_1 - \epsilon_2\right) 
\end{pmatrix}
\end{align}
\end{widetext}
and the positive definite matrix
\begin{align}
D= 
\begin{pmatrix}
\zeta^2 \gamma_1^+   & 0
\\
0  & \zeta^2 \gamma_2^+
\end{pmatrix}
\end{align}
the dynamics of the covariance matrix $C$ reads
\begin{align}
\frac{d C}{dt} = WC + CW^{\dagger} + D.
\label{eq:ct}
\end{align}
The thermodynamic observables, such as heat currents $J_1(t)$ and $J_2(t)$, are related with the elements of the covariance matrix $C$ as follows,
\begin{align}
J_{1} = & \zeta^2 \left\lbrace  -\frac{\gamma_1^+ + \gamma_1^- }{2}\left[  4\epsilon_1 C_{11} + \lambda\left(  C_{12} + C_{21}\right) \right] 
+2\epsilon_1\gamma_1^+\right\rbrace  ,
\\
J_{2} = & \zeta^2 \left\lbrace  -\frac{\gamma_2^+ + \gamma_2^- }{2}\left[ 4\epsilon_2  C_{22} + \lambda\left( C_{12} + C_{21}\right) \right]  +2\epsilon_2\gamma_2^+\right\rbrace.
\end{align}

In order to investigate the characteristics of the system in its steady state, one only needs to solve the following Lyapunov equation,
\begin{align}
	WC + CW^{\dagger} + D = 0,
\end{align}
and it is not difficult to find the algorithms for solving this type of equation in the scientific computing libraries. 

Regarding the study of the system's transient behavior, it is necessary to examine the analytical solution of Eq.~(\ref{eq:ct}), which is given as follows \cite{RevModPhys.94.045006},
\begin{align}
C(t) = e^{Wt} \left[ C(0) + \int_{0}^{1} dt' e^{-Wt'} D e^{-W^{\dagger}t'}\right] e^{W^{\dagger}t}.
\end{align}
The convergence of $C(t)$ is ensured as long as the real parts of all eigenvalues of the matrix $W$ are negative. Furthermore, if the eigenvalues of $W$ are non-degenerate, the corresponding eigenvectors constitute a complete basis, enabling the expansion of other matrices, such as $C(0)$ and $D$, in terms of this set of basis. Under our parameter setting specified in the main text, both these two conditions can be met simultaneously. In this case, the relaxation time $\tau_r$ of $C(t)$ can be obtained by analyzing the eigenvalues of $W$.

Denoting the eigenvalue equation of $W$ by
\begin{align}
W\ket{\mathcal{W}_i} = w_i\ket{\mathcal{W}_i}, (i = 1,2),
\end{align}
the matrices $C(0)$ and $D$ can be decomposed as
\begin{align}
& C(0) = \sum_{i,j} c_{ij}(0) \ket{\mathcal{W}_i} \bra{\mathcal{W}_j},
\\
& D = \sum_{i,j} d_{ij} \ket{\mathcal{W}_i} \bra{\mathcal{W}_j},
\end{align}
and $C(t)$ can be rewritten into
\begin{align}
C(t)  =& \sum_{i,j} c_{ij}(0) e^{\left( w_i + w_j^*\right) t}\ket{\mathcal{W}_i} \bra{\mathcal{W}_j}
\notag
\\
& + \sum_{i,j} \frac{\left[ 1 - e^{-\left( w_i + w_j^*\right)} \right] d_{ij} }{ w_i + w_j^*} e^{\left( w_i + w_j^*\right) t}\ket{\mathcal{W}_i} \bra{\mathcal{W}_j}.
\end{align}
This indicates that $C(t)$ can be expressed as the linear combination of a series of exponentially decay terms. Introducing the parameter
\begin{align}
v \equiv \left| 2\max \left( \Re w_i \right) \right|,
\end{align}
where $\Re w_i$ represents the real part of $w_i$ and $\max$ stands for the maximum operation, then it is reasonable to define $\tau_r$ as
\begin{align}
\tau_r = v^{-1}.
\end{align}
The role of $\tau_r$ is to roughly characterize the time period required for the system to attain a steady state. When the system's evolution time $t$ is a multiple of $\tau_r$, it can be approximately considered that the system is nearing the steady state. Generally, $\tau_r$ is a complex function that depends on other parameters. However, in the main text, we have chosen the parameters such that $|\gamma^+_1 + \gamma^-_1 - (\gamma^+_2 + \gamma^-_2)| \sim 1$ and both $\lambda$ and $\zeta^2$ are much smaller than  $\epsilon_1 - \epsilon_2$. Given this choice of the parameters, it is straightforward to demonstrate that $\Re w_i \propto \zeta^2$ approximately holds true. Consequently, according to the definition of $\tau_r$, we conclude that $\tau_r \propto \zeta^{-2}$.
}

\bibliography{mybib}

\end{document}